\documentclass{mn2e}
\usepackage{epsfig}
\usepackage{epsf}

\title[Spitzcar]{Spitzer Space Telescope observations of the Carina
  Nebula: The steady march of feedback-driven star
  formation\thanks{Based on observations made with the Spitzer Space
    Telescope, which is operated by the Jet Propulsion Laboratory,
    California Institute of Technology under a contract with
    NASA. Support for this work was provided by NASA through an award
    issued by JPL/Caltech.}}

\author[N.\ Smith et al.]{Nathan Smith$^1$\thanks{Email:
    nathans@astro.berkeley.edu}, Matthew S.\ Povich$^{2,3}$\thanks{NSF
    Astronomy \& Astrophysics Fellow.}, Barbara A.\ Whitney$^4$, Ed
  Churchwell$^2$, \newauthor Brian L.\ Babler$^2$, Marilyn R.\
  Meade$^2$, John Bally$^5$, Robert D.\ Gehrz$^6$, \newauthor Thomas
  P.\ Robitaille$^7$\thanks{Spitzer Postdoctoral Fellow.}, \& Keivan
  G.\ Stassun$^8$ \\ $^1$Astronomy Department, University of
  California, 601 Campbell Hall, Berkeley, CA 94720, USA \\
  $^2$Astronomy Department, University of Wisconsin, 475 N. Charter
  Street, Madison, WI, 53706, USA \\ $^3$Department of Astronomy \&
  Astrophysics, Pennsylvania State University, 525 Davey Laboratory,
  University Park, PA 16801, USA \\ $^4$Space Science Institute, 4750
  Walnut Street, Suite 205, Boulder, CO 80301, USA \\ $^5$Center for
  Astrophysics and Space Astronomy, University of Colorado, 389 UCB,
  Boulder, CO 80309, USA \\ $^6$Astronomy Department, University of
  Minnesota, 116 Church St. SE, Minneapolis, MN, 55454, USA \\
  $^7$Harvard-Smithsonian Center for Astrophysics, 60 Garden Street,
  Cambridge, MA 02138, USA \\ $^8$Physics \& Astronomy Department,
  Vanderbilt University, 1807 Station B, Nashville, TN, 37235, USA}

\begin{document}
\date{Accepted 0000, Received 0000, in original form 0000}
\pagerange{\pageref{firstpage}--\pageref{lastpage}} \pubyear{2009}
\def\arcdeg{\degr}
\maketitle
\label{firstpage}

\begin{abstract}

  We report the first results of imaging the Carina Nebula (NGC~3372)
  with the Infrared Array Camera (IRAC) onboard the {\it Spitzer Space
    Telescope}, providing a photometry catalog of over 44,000 point
  sources as well as a catalog of over 900 candidate young stellar
  objects (YSOs) based on fits to their spectral energy distributions
  (SEDs).  We discuss several aspects of the extended emission,
  including the structure of dozens of dust pillars that result when a
  clumpy molecular cloud is shredded by feedback from massive stars.
  There are surprisingly few of the ``extended green objects'' (EGOs)
  that are normally taken as signposts of outflow activity in {\it
    Spitzer} data, and not one of the dozens of Herbig-Haro jets
  detected optically are seen as EGOs.  EGOs are apparently poor
  tracers of outflow activity in strongly irradiated environments, due
  to the effects of massive star feedback.  A population of ``extended
  red objects'' tends to be found around late O-type and early B-type
  stars, some with clear bow-shock morphology.  These are dusty shocks
  where stellar winds collide with photoevaporative flows off nearby
  clouds.  Finally, the relative distributions of O-type stars, small
  star clusters, and sub-clusters of YSOs as compared to the dust
  pillars shows that while some YSOs are located within dust pillars,
  many more stars and YSOs reside just outside pillar heads. We
  suggest that pillars are transient phenomena, part of a continuous
  outwardly propagating wave of star formation driven by feedback from
  massive stars.  As the pillars are destroyed, they leave newly
  formed stars in their wake, and these are then subsumed into the
  young OB association.  The YSOs are found predominantly in the
  cavity between pillars and massive stars, arguing that their
  formation was in fact triggered.  Altogether, the current generation
  of YSOs shows no strong deviation from a normal initial mass
  function (IMF).  The number of YSOs is consistent with a roughly
  constant star-formation rate over the past $\sim$3 Myr, implying
  that propagating star formation in pillars constitutes an important
  mechanism to construct unbound OB associations.  These accelerated
  pillars may give birth to massive O-type stars that, after several
  Myr, could appear to have formed in isolation.

\end{abstract}

\begin{keywords}
  H~{\sc ii} regions --- ISM: individual (NGC~3372) —-- stars:
  formation --— stars: luminosity function, mass function --- stars:
  pre-main-sequence
\end{keywords}

%%%%%%%%%%%%%%%%%%%%%%%%%%%%%%%%%%%%%%%%%%%%%%%%%%%%%%%%%%%%%%%%%%%%%%%%%%
\section{INTRODUCTION}

Most stars form in OB associations, where the stellar winds, UV
radiation, and eventual supernova (SN) explosions from the massive
members significantly impact the environment.  The feedback from newly
born massive stars will ultimately destroy their natal molecular
cloud, clearing away the dust and gas and thereby shutting off star
formation and determining the star formation efficiency.  In the
process, the same feedback may simultaneously trigger the birth of new
generations of stars, allowing star formation to propagate
continuously from one point to the next (Elmegreen \& Lada 1977).
This distributes the star formation in space and time, giving rise to
significant age spreads in the resulting stars.  This mode of star
formation may give rise to large OB associations, rather than compact
single star clusters.  The southern Milky Way provides a striking
example of this contrast: the Carina Nebula and the cluster NGC~3603
have roughly the same number of O-type stars and roughly the same
ionizing photon output (Smith 2006; Crowther \& Dessart 1998), but
NGC~3603 is dominated by a single dense cluster within a radius of
$\sim$1 pc, whereas the O-type stars in Carina are distributed in
several sub-clusters over 10--20 pc.

Evidence suggests that the dust pillars commonly seen at the borders
of H~{\sc ii} regions may be prime sites for this mode of propagating
star formation (Bally \& Reipurth 2003; McCaughrean \& Andersen 2002;
Jiang et al.\ 2002; Stanke et al.\ 2002; Smith et al.\ 2005; Rathborne
et al.\ 2004).  Whether or not the second generation of stars forming
in the pillars was triggered is still unclear: it is difficult to
verify whether the stars began to form spontaneously due to initial
clumps in the cloud and were simply uncovered by the advancing
ionization front, whether their parent clumps were agglomerated in the
``collect and collapse'' scenario (Elmegreen 1992) where these clumps
subsequently formed stars under their own self-gravity, or if they
were triggered directly via radiation-driven implosion (Oort \&
Spitzer 1955; Kahn 1969; Dyson 1973; Elmegreen 1976; Bertoldi 1989;
Bertoldi \& McKee 1990; Williams et al.\ 2001; Gorti \& Hollenbach
2002).  Star formation in these environments plays a critical role in
the evolution of H~{\sc ii} regions and the interstellar medium (ISM)
(e.g., Elmegreen \& Scalo 2004).

The Carina Nebula is a special case among massive star-forming regions
in the Milky Way: like the Orion Nebula, it is near enough (2.3 kpc;
Smith 2006) to facilitiate studies of the details of faint nebular
emission and the population of low to intermediate mass stars forming
alongside massive stars, but unlike Orion, Carina has over 70 O-type
stars and samples the top end of the stellar mass function.  This
provides a large region over which to investigate the impact of
massive-star feedback.  The global properties of the Carina Nebula
have been discussed recently by Smith \& Brooks (2007).  See Smith
(2006) for a census of the massive stars that power Carina, and see
Walborn (1995, 2009) for excellent reviews of the remarkable
collection of extreme O-type stars in the region.  The stellar content
of the massive central clusters Trumpler (Tr) 14 and 16 includes
$\eta$~Carinae, three WNH stars, and a number of the most extreme
early O-type stars known (e.g., Walborn et al.\ 2002a).  In fact,
Carina is the first region where the earliest spectral type O stars
were recognized (Walborn 1973).  All other regions in the Milky Way
with a comparable stellar population are more distant and more
obscured by dust.  Thus, Carina provides a laboratory for low-mass
star formation and protoplanetary disks in regions analogous to much
more extreme starbursts that are far too distant for such studies.

%%%%%%%%%%%%%%%%%%%%%%%%% FIGURE 1 - MAP  %%%%%%%%%%%%%%%%%%%%%%%%%%%
\begin{figure}\begin{center}
%\epsscale{0.99}
%\includegraphics[width=3.1in]{../MAP/msxmap.eps}
\includegraphics[width=3.1in]{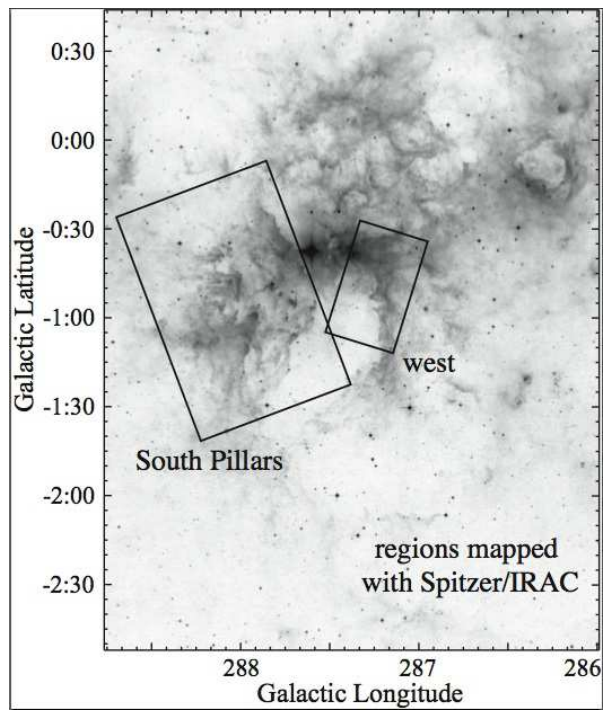}
\end{center}
\caption{A large-scale 8.6~$\mu$m image of the Carina Nebula obtained
  with {\it MSX}, from Smith \& Brooks (2007).  The regions mapped
  with IRAC are shown.  The clusters Tr~14 and Tr~16 are located
  between these two fields amid the brightest diffuse
  emission.}\label{fig:map}
\end{figure}
%%%%%%%%%%%%%%%%%%%%%%%%%%%%%%%%%%%%%%%%%%%%%%%%%%%%%%%%%%%%%%%%%%%%%%

%%%%%%%%%%%%%%%%%%%%%%%%% FIGURE 2 - color IRAC  %%%%%%%%%%%%%%%%%%%%
\begin{figure*}\begin{center}
%\epsscale{0.99}
%\includegraphics[width=6.7in]{../FIGS/spitzerSP.eps}
\includegraphics[width=6.7in]{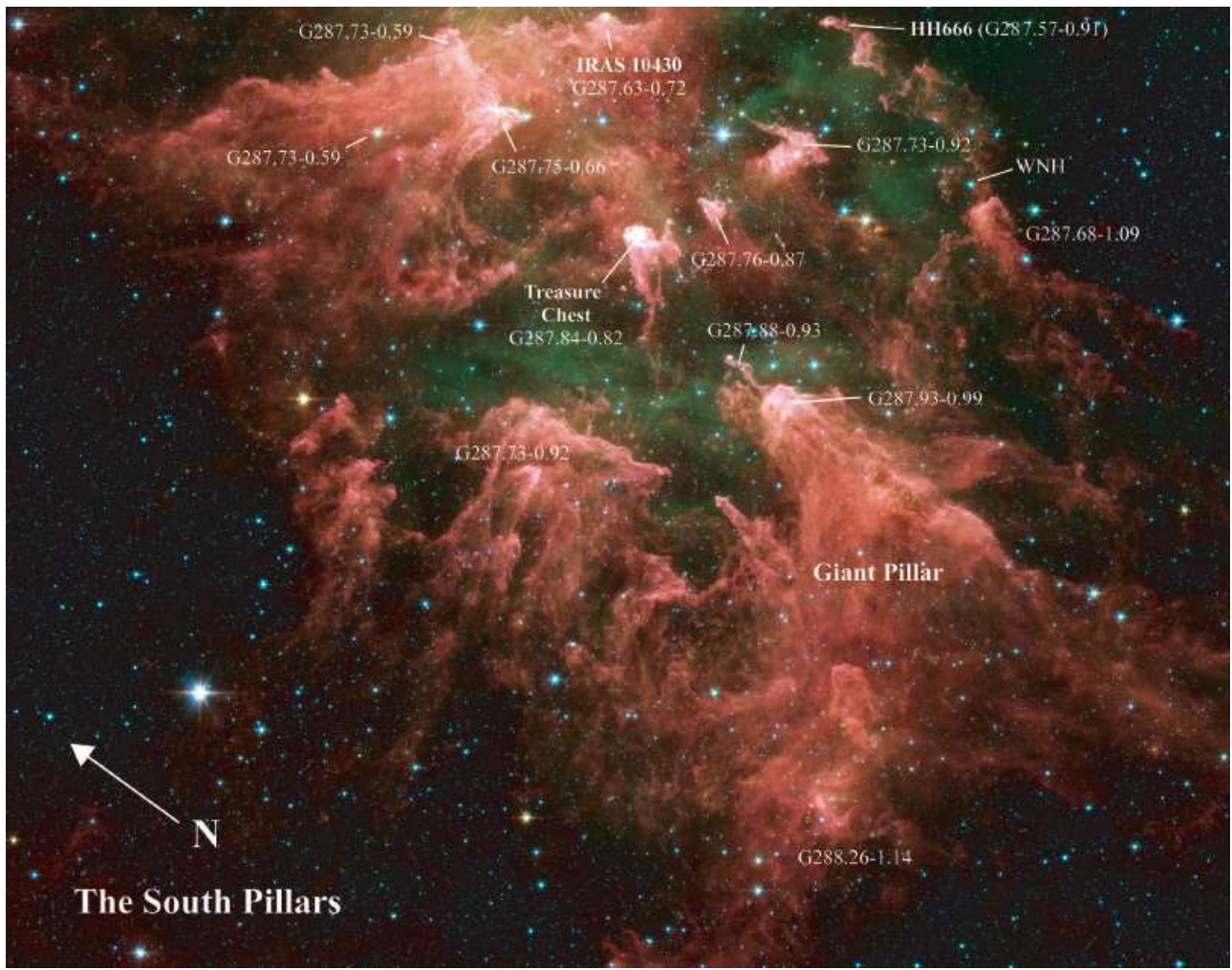}
\end{center}
\caption{Color image of the South Pillars made from IRAC data with 3.6
  $\mu$m (blue), 4.5 $\mu$m (green), 5.8 $\mu$m (orange), and 8.0
  $\mu$m (red).  Several previously studied features are labeled.  The
  size scale and field of this image are shown in
  Figure~\ref{fig:map}.  Tr~14 and Tr~16 are located off the top of
  this page.  The center of this field is at roughly 10$^{\rm h}$47$^{\rm m}$, 
  --60$\arcdeg$05$\arcmin$ (J2000). }\label{fig:colorSP}
\end{figure*}
%%%%%%%%%%%%%%%%%%%%%%%%%%%%%%%%%%%%%%%%%%%%%%%%%%%%%%%%%%%%%%%%%%%%%%

Although a great number of massive stars have formed in the past
$\sim$3 Myr in Carina, the current hot-bed of star formation activity
has migrated to the southern part of the nebula, in a region called
the South Pillars (Smith et al.\ 2000).  The widespread ongoing
star-formation activity in the South Pillars, as well as the structure
of the pillars themselves, was first recognized based on wide-field IR
images (Smith et al.\ 2000) obtained with the {\it Midcourse Space
  Experiment} ({\it MSX}).  (There appear to be several other regions
of ongoing star formation to the north and west as well, which are
less active than the South Pillars.)  An 8.6 $\mu$m {\it MSX} image of
the Carina Nebula is shown in Figure~\ref{fig:map}.  The presence of
active, ongoing star formation in this region has since been confirmed
by obervations of embedded young clusters like the ``Treasure Chest''
and others (Smith et al.\ 2005; Rathborne et al.\ 2002, 2004; H\"agele
et al.\ 2004) and a large number of outflows in the form of irradiated
Herbig-Haro (HH) jets throughout the region (Smith et al.\ 2004a,
2010).  There are many dense cores detected in C$^{18}$O that are
spread across the region, providing sites of ongoing and potential
future star formation, with some evidence for molecular outflow
activity (Yonekura et al.\ 2005).

The South Pillars are intriguing because they provide a powerful
laboratory in which to study the details of the feedback mechanism by
which massive stars destroy their natal molecular clouds.  In the
process, the massive stars influence subsequent generations of stars
and planets forming in the surrounding region.  Perhaps the birth of
some of those newly formed stars was even directly triggered by the
external feedback.  The massive stars also clear away and illuminate
the surrounding gas as we witness the assemblage of an OB association.
Study of regions like the South Pillars, where young stars are forming
in the immediate vicinity of several dozen stars that will explode as
supernovae in the next 1--2 Myrs, may have implications for our own
Solar System (see Smith \& Brooks 2007).

In this paper we present the first results from observations of the
Carina Nebula obtained with the {\it Spitzer Space Telescope} (Werner
et al.\ 2004; Gehrz et al.\ 2007).  This is the first systematic study
of the IR properties of the stars, dust, and gas in the South Pillars
of Carina, so our aim here is to give a broad overview of the region
as seen at 3-8 $\mu$m; more detailed multi-wavelength studies will
follow.  We present our observations in \S 2, we discusss the
photometry of point sources and the distribution of observed
properties among the point source spectral energy distributions (SEDs)
in \S 3, and in \S 4 we discuss various properties of the extended gas
and dust emission and how it relates to nearby stars.  This includes
extended green objects (\S 4.2), extended red objects (\S 4.3),
relationships between dust pillars and massive stars (\S 4.4 and 4.5),
and the discussion of a newly recognized star cluster near Tr~16 that
was obscured behind a dark dust lane at visual wavelengths (\S 4.6).
In \S 5 we synthesize the main results, and in \S 6 we recap with a
list of our main conclusions.  There are several additional and more
detailed results and implications from these data, which cannot be
described in full here, but will be included in future papers.

%%%%%%%%%%%%%%%%%%%%%%%%% FIGURE 3 - color west  %%%%%%%%%%%%%%%%%%%
\begin{figure}\begin{center}
%\epsscale{0.99}
%\includegraphics[width=3.0in]{../FIGS/spitzerWEST.eps}
\includegraphics[width=3.0in]{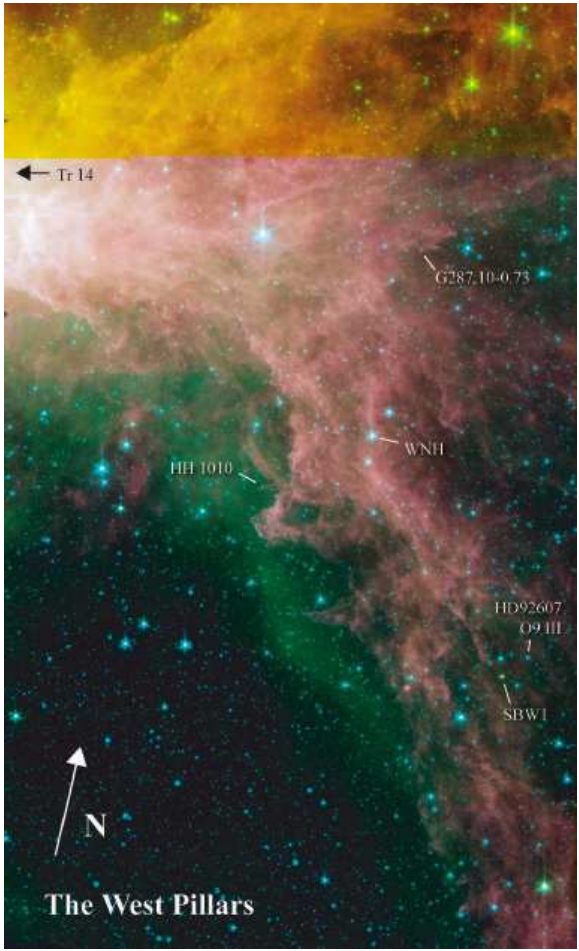}
\end{center}
\caption{Same as Figure~\ref{fig:colorSP}, but showing the western
  field with an adjusted intensity scale.  The greenish/red region at
  the top was at the edge of the field and was observed in Bands 2
  (4.5 $\mu$m) and 4 (8.0 $\mu$m), but not 1 (3.6 $\mu$m) and 2 (5.8
  $\mu$m).  We include it here because it contains interesting
  structure near Tr~14.  The orientation and field of view are shown
  in Figure~\ref{fig:map}.}\label{fig:colorWest}
\end{figure}
%%%%%%%%%%%%%%%%%%%%%%%%%%%%%%%%%%%%%%%%%%%%%%%%%%%%%%%%%%%%%%%%%%%%%%
%%
%%%%%%%%%%%%%%%%%%%%%%%%%%%%%%%%%%%%%%%%%%%%%%
\section{OBSERVATIONS}

We used the Infrared Array Camera (IRAC; Fazio et al.\ 2004) 
onboard {\it Spitzer} to map portions of the Carina Nebula in the four
bandpasses available with that instrument, centered at 3.6, 4.5, 5.8,
and 8.0 $\mu$m.  Because of its intrinsic expanse and its proximity,
diffuse emission associated with the Carina Nebula covers many square
degrees on the sky at optical to mid-IR wavelengths (see Smith \&
Brooks 2007).  The two fields we targeted with IRAC, shown in
Figure~\ref{fig:map} superposed on an image obtained with the {\it
  Midcourse Space Experiment} ({\it MSX}; Egan et al.\ 1998), were
chosen based on earlier results from {\it MSX} showing extended mid-IR
emission at lower resolution in Carina (Smith et al.\ 2000; Smith \&
Brooks 2007).  Motivated by these earlier resuts, we chose to observe
the South Pillar region because it represented the complex structure
of a clumpy molecular cloud being shredded by feedback from nearby
massive stars, resulting in an army of dust pillars with signs of
current embedded star formation amid a large number of recently formed
young stars.  The western region was chosen because it represented a
clean edge-on view of an ionization front and photodissiciation
region, also showing signs of active ongoing star formation.  In both
regions, dust pillars point back to the central part of the nebula,
suggesting that they are at the same distance and that the dust
pillars are in fact shaped by those massive stars (see Smith et al.\
2000).

We did not target central regions of the nebula near $\eta$~Carinae
itself using IRAC because of potential danger to the IRAC detectors.
% (diffraction spikes from $\eta$ Car can be seen at the north-western
% side of the IRAC images of the South Pillars [top of
% Figure~\ref{fig:colorSP}], the edge of which came close to $\eta$
% Car).
{\it MSX} also showed possible evidence for ongoing star formation in
regions of Carina that are north of Tr~14 and Tr~16, but these regions
seemed to have less vigorous star formation activity, and interpreting
their true location was complicated by geometrical effects and the
presence of the Tr~15 cluster, whose distance is not as certain as
Tr~14 and Tr~16.  Therefore, we did not target regions to the north in
this program.

The IRAC data were reduced and combined into a large mosaic using the
pipeline developed for the Galactic Legacy Infrared Midplane Survey
Extraordinaire (GLIMPSE; see Benjamin et al.\ 2003; Churchwell et al.\
2009).\footnote{See {\tt
    http://www.astro.wisc.edu/glimpse/docs.html}.}  With the aim of
studying the low-mass (1--3 $M_{\odot}$) young stellar population, we
obtained deeper images at each position (16 $\times$ 2 sec) than were
obtained at most positions in the GLIMPSE survey (2 $\times$ 2 sec)
where maximizing areal coverage was the driving factor.  Additionally,
the South Pillars are $\sim$1\arcdeg\ out of the Galactic plane, so
the confusion limit was less restrictive than in most parts of the
inner Galactic plane seen by GLIMPSE.  The final mosic pixel scale was
0$\farcs$6, subsampled by 2 from the raw pixel scale of 1$\farcs$2,
with a FWHM spatial resolution of roughly 2\arcsec\ at IRAC
wavelengths.  The full mosaic color images of the South Pillars and
the western region made with IRAC are shown in
Figures~\ref{fig:colorSP} and \ref{fig:colorWest}, respectively.

To create a catalog of point source photometry from the IRAC data, we
used the GLIMPSE point-source extractor tool, which is a modified
version of DAOPHOT.  This resulted in the detection of 53,905 point
sources, 48,642 of which were in the area observed by all four IRAC
bands (i.e., there were 5,263 sources that fell at the ends of the
mapped regions and were only detected in Bands 1 and 3, or Bands 2 and
4). We then cross-referenced these IRAC point sources with the
Two-Micron All Sky Survey (2MASS) point source catalog (Skrutskie et
al.\ 2006) to produce a merged 2MASS+IRAC Carina point-source catalog
of 44,032 sources that were detected in four or more filters (36,720
in the South Pillars field, and 7312 in the western region; see
Figure~\ref{fig:map}).  Of these, 42,626 were fit by stellar
photospheres, many of which are foreground and background sources.
Carina is seen along the tangent point of the Carina spiral arm,
providing substantial background contamination.  1,406 sources could
not be well fit by photospheres, and many of these are young stellar
objects (YSOs) as discussed in the following section.  A brief summary
of point source results is given in Table~\ref{tab:psf}, and an
electronic version of the Carina point source catalog is available at
{\tt http://www.astro.wisc.edu/glimpse/Carina\_Nebula/}.

%%%%%%%%%%%%%%%%%%%%%%%%% FIGURE 4 - YSOs on image  %%%%%%%%%%%%%%%%%%
\begin{figure*}\begin{center}
%\epsscale{0.99}
%\includegraphics[width=5.6in]{../Povich/gray_filters.eps}
\includegraphics[width=5.6in]{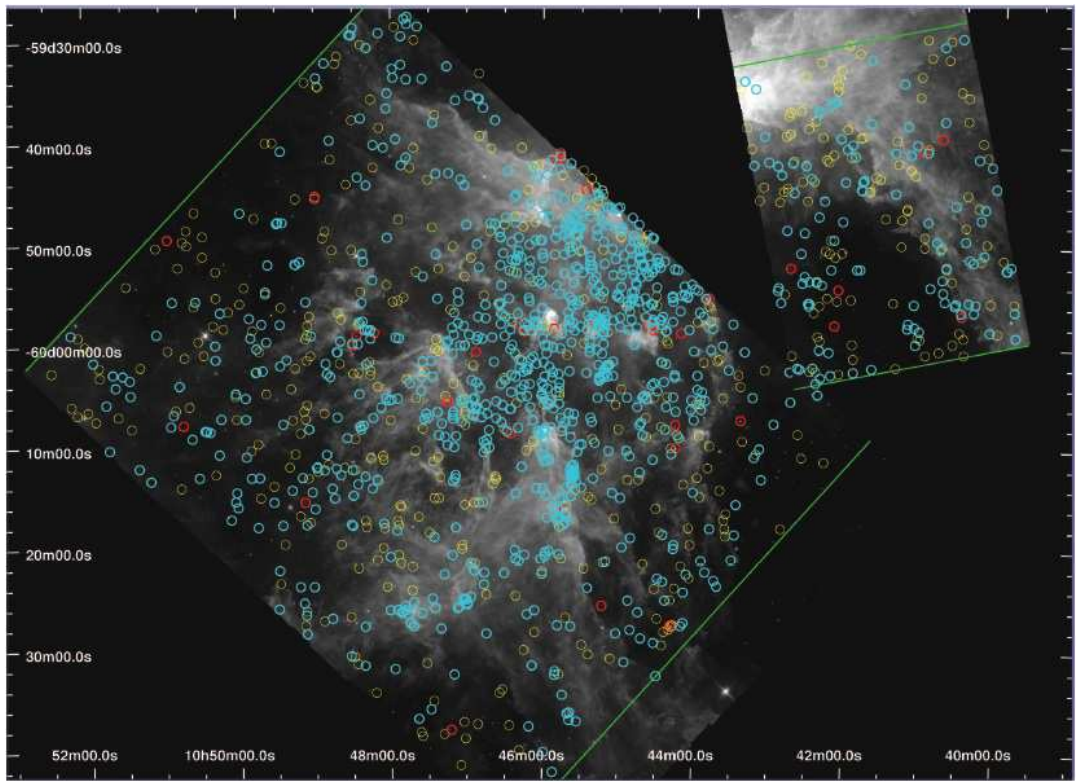}
\end{center}
\caption{Spitzer/IRAC Band 4 (8.0~$\mu$m) image of the south pillars
  and the west region in greyscale, superposed with small circles
  noting the positions of YSO sources.  Cyan circles are YSOs that met
  our criteria for reliable IR excess and were well-fit by YSO models,
  yellow circles were poorly fit by stellar photospheres but did not
  meet our criteria for agreeing with YSO models, and red circles
  passed our criteria explained in the text for IR excess but were not
  well-fit by existing YSO SEDs.}\label{fig:ysos}
\end{figure*}
%%%%%%%%%%%%%%%%%%%%%%%%%%%%%%%%%%%%%%%%%%%%%%%%%%%%%%%%%%%%%%%%%%%%%%

%%%%%%%%%%%%%%% TABLE 1 - point source stats  %%%%%%%%%%%%%%%%%%%%%%%%%
\begin{table*}\begin{minipage}{5.5in}
\caption{IRAC Point Source Summary}\scriptsize
\begin{tabular}{@{}lrrrl}\hline\hline
Parameter  &South Pillars &West Region &Total &Comment\\ \hline
%%-------------------------------------------------------------
Total        &44,428 &9,477 &53,905 &all detected IRAC point sources  \\
Cropped      &40,605 &8,037 &48,642 &in region observed by 4 IRAC bands  \\
2MASS+IRAC   &36,720 &7,312 &44,032 &merged, detected in at least 4/7 filters  \\
Photospheres &35,531 &7,095 &42,626 &well-fit by stellar photospheres  \\
Excess       &1,189  &217   &1,406  &poorly fit by stellar photospheres  \\

candidate YSOs&804   &105  &909  &SED well fit by YSO models  \\
Stage 0/I    &140    &17   &157  &SED well fit by YSO models   \\
Stage II     &195    &19   &214  &SED well fit by YSO models   \\
Stage III    &37     &11   &48   &SED well fit by YSO models   \\
Ambiguous    &432    &58   &490  &SED well fit by YSO models   \\
Class I      &151    &13   &164  &color-color \\
Class II     &451    &63   &514  &color-color \\
Class III    &140    &17   &157  &color-color \\
\hline
\end{tabular}\label{tab:psf}
\end{minipage}
\end{table*}

%%%%%%%%%%%%%%%%%%%%%%%%% FIGURE 5 - YSO class on image  %%%%%%%%%%%%%%%%%%
\begin{figure*}\begin{center}
%\epsscale{0.99}
%\includegraphics[width=5.6in]{../Povich/gray_stages.eps}
\includegraphics[width=5.6in]{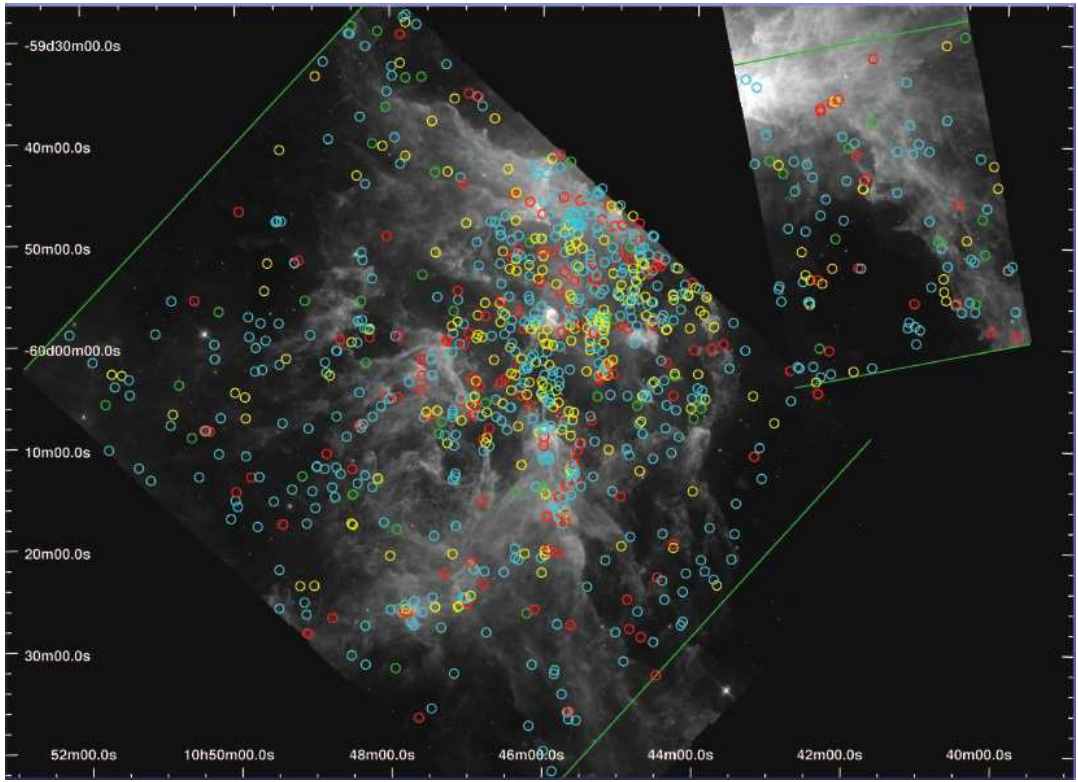}
\end{center}
\caption{Same as Figure~\ref{fig:ysos}, but plotting the locations of
  only the well fit YSOs (those that are cyan in
  Figure~\ref{fig:ysos}), color-coded by evolutionary stage for each
  object as defined in the text (note that evolutionary ``stage'' as
  defined here is not identical to the usual Class I, II, III, etc.,
  for reasons having to do with how they were classified by the fitter
  routine; see text).  Cyan circles are those with ambiguous stages,
  retaining their cyan color. Red circles are Stage 0/I, yellow
  circles are Stage II, and green circles are Stage
  III.}\label{fig:stage}
\end{figure*}
%%%%%%%%%%%%%%%%%%%%%%%%%%%%%%%%%%%%%%%%%%%%%%%%%%%%%%%%%%%%%%%%%%%%%%
%
\section{Point Sources}
%%%%%%%%%%%%%%%%%%%%%%%%%%%%%%%%%%%%%%%%%%%%%%%%%%%%%%%%%%%
\subsection{SED-Based Selection and Analysis of YSOs}

The South Pillar and Western regions we observed with {\it
  Spitzer}/IRAC include many thousands of point sources projected on
the sky amid dust pillars and other diffuse emission, but the ones
that we are most interested in are the young stellar objects (YSOs)
that trace the current and recent star formation in the region.  Some
are still embedded in the pillars, and many more are not.  With the
point source catalog derived from the {\it Spitzer} observations of
the South Pillars and the West region in Carina, we used the grid of
young stellar object (YSO) models from Robitaille et al.\ (2006) and
the SED fitting tool of Robitaille et al.\ (2007) to identify and
characterize candidate YSOs.

The model grid consists of 20,000 two-dimensional Monte Carlo
radiation transfer models (Whitney et al.\ 2003a, 2003b, 2004)
spanning a complete range of stellar mass and evolutionary stage and
output at 10 viewing angles (inclinations), so the fitting tool has
200,000 SEDs to choose among when applied to the broadband photometric
data. An additional set of 7853 model stellar atmospheres and the
mid-IR interstellar extinction law of Indebetouw et al.\ (2005) were
included in the Robitaille et al.\ (2007) fitting tool to facilitate
filtering of sources consistent with stellar photospheres reddened by
interstellar dust only.  The procedure we used to select candidate
YSOs was described in detail by Povich et al.\ (2009; see also
Shepherd et al.\ 2007; Indebetouw et al.\ 2007).

To summarize, we first fit stellar atmosphere SEDs to 44,032 sources
from the highly-reliable point-source catalogs (including both the
South Pillars and West fields) that met the following criteria: (1)
Sources located in regions of the sky observed in all 4 IRAC bands;
and (2) Sources detected in $N_{\rm data}\ge 4$ out of the 7 combined
2MASS+IRAC bands in the catalog.  To avoid biasing the fits for
sources where the photometric uncertainties may have been
underestimated in the catalog, we conservatively reset the
uncertainties to a minimum value of 10\% before fitting any models.
We considered a source to be well-fit by a stellar atmosphere if the
(non-reduced) goodness-of-fit parameter, normalized by the number of
distinct bands in which the source was detected, satisfied
$\chi^2/N_{\rm data} \le 2$. The number of sources thus identified as
normally reddened stellar photospheres was 42,626. The remaining 1406
sources, shown in Figure~\ref{fig:ysos}, could not be fit well by
stellar atmospheres because either (1) they exhibit IR emission
circumstellar dust that creates an excess emission above the stellar
photosphere, or (2) they suffer from systematic photometric
uncertainties larger than the reset uncertainties from the catalog,
which can create spurious IR excess emission.

The most common case of spurious IR excess emission occurs when a star
is detected in the IRAC 3.6 and 4.5~$\mu$m bands and its flux density
is significantly over-estimated in the IRAC 8.0 or (less commonly)
5.8~$\mu$m band because a noise peak or diffuse emission is extracted
at the position of the point source (Povich et al.\ 2009). This most
often occurs for faint sources near the detection limit, especially in
regions of bright diffuse emission, since stars are fainter at longer
wavelengths, diffuse emission is brighter, and the IRAC 5.8 and
8.0~$\mu$m bands have lower sensitivity. For this work, we have
developed a series of color cuts to automatically cull out sources
with probable spurious IR excess, where ``spurious'' is defined as:

\begin{enumerate}
\item {\em [8.0] excess only.} IR excess occurs only at 8.0~$\mu$m;
  all other bands consistent with a stellar photosphere.\footnote{Note
    that this will automatically cut out any ``transition disks'' that
    may be present in the sample.  Because of the bright background in
    Band 4, the inclusion of data at longer wavelengths is necessary
    to study these sources with possible inner disk holes.}
\item {\em [5.8] excess only.} IR excess occurs only at 5.8~$\mu$m;
  all other bands consistent with a stellar photosphere (in this case
  the source is almost never detected at 8.0~$\mu$m).
\item {\em [5.8] and [8.0] displacement.} IR excess at both 5.8 and
  8.0~$\mu$m, but the color $[5.8]-[8.0] = 0$ as in a stellar
  photosphere. This is relatively uncommon among Catalog sources, and
  the explanation is not clear.
\end{enumerate}

To obtain a highly-reliable sample of candidate YSOs, we begin by
selecting only sources detected at both 3.6 and 4.5~$\mu$m. Sources
are then considered candidate YSOs if the following conditions are
met:
\begin{eqnarray}
   [3.6]-[4.5] & > & \delta([3.6]-[4.5]); \\
{\rm OR}~~ |[4.5]-[5.8]| & > & \delta([4.5]-[5.8]), \\
   {[5.8]-[8.0]} & > & \delta([5.8]-[8.0]);
\end{eqnarray}
where the $[\lambda]$ denote magnitudes and the
$\delta([\lambda_i]-[\lambda_j])$ are uncertainties on the colors
computed from the (minimum 10\%) uncertainties on catalog flux
densities. 454/1406 sources poorly-fit by stellar photospheres fail to
meet the conditions of Equations 1--3; these are plotted as {\em
  yellow} circles in Figure~\ref{fig:ysos}.

Finally, we fit the 952 sources that have passed through all the above
filters with YSO models, and apply the same $\chi^2/N_{\rm data} \le
2$ criterion to classify sources as well-fit by YSO SEDs. The large
majority, 909/953, are well-fit by YSO models, and these are plotted
as {\em cyan} circles in Figure~\ref{fig:ysos}; these sources are
entered into our YSO catalog (Table Y1). The remaining 43 sources are
plotted as {\em red} circles.

The YSO models can be divided into evolutionary stages that parallel
the traditional observational ``class'' taxonomy, where Stage 0/I are
deeply embedded protostars, Stage II are pre-main sequence (PMS) stars
with optically thick circumstellar disks, and Stage III are PMS stars
with optically thin disks (see section 3.1 of Robitaille et al.\
2006).  In general, a given source can be fit by multiple YSO models,
so for each YSO in the catalog we define a set of well-fit models
according to
\begin{equation}\label{best}
  \frac{\chi^2}{N_{\rm data}} - \frac{\chi^2_{\rm min}}{N_{\rm data}} \le 2,
\end{equation}
where $\chi^2_{\rm min}$ is the goodness-of-fit parameter for the
best-fit model (Povich et al.\ 2009). If $\ge 0.67$ of the models
(weighted by $\chi^2$) fit to a given source correspond to a single
evolutionary stage, then we assign that stage to that source,
otherwise the stage is considered {\it ambiguous}. This is
philosophically different from the approach of classifying {\em all}
candidate YSOs on the basis of colors or spectral indices (e.g.\ Allen
et al.\ 2004) because it acknowledges the model-dependent nature of
the physical interpretation of the classifications and the blurry
boundaries between the classes. It also acknowledges degeneracies in
color and evolutionary stage. Locations of the 909 sources well-fit by
YSO models are plotted in Figure~\ref{fig:stage}, color-coded
according to evolutionary stage. Note that 54\% (490/909) of the
candidate YSOs plotted in Figure~\ref{fig:stage} have ambiguous stage
classifications, and retain their {\it cyan} colors from
Figure~\ref{fig:ysos}. This relatively high fraction of ambiguous
sources could be reduced with additional constraints on the SEDs at
$\lambda > 10~\mu$m; MIPS 24~$\mu$m photometry, for example, has
proved useful to help discriminate between disk-dominated and
envelope-dominated YSOs (Indebetouw et al.\ 2007; Povich et al.\
2009).  A summary of YSO properties is provided in Table 1, although
note that the numbers of sources given in Table 1 are certainly
underestimates for the true number of YSOs and their subtypes (see \S
3.3).

%%%%%%%%%%%%%%%%%%%%%%%%% FIGURE 6 - stages  %%%%%%%%%%%%%%%%%%
\begin{figure}\begin{center}
%\epsscale{0.99}
%\includegraphics[width=3.2in]{../Povich/stages.eps}
\includegraphics[width=3.2in]{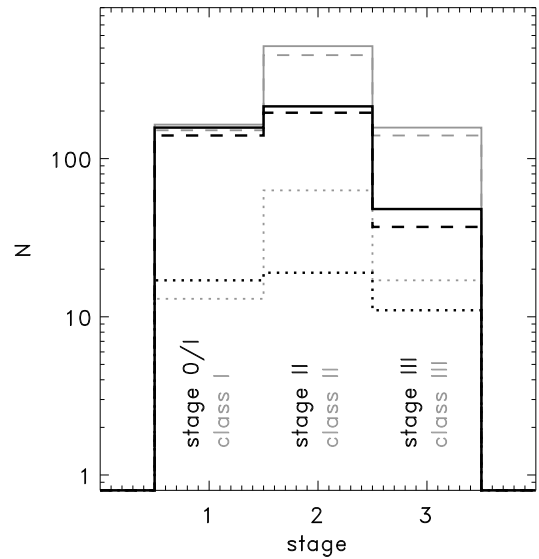}
\end{center}
\caption{Histogram of YSO stages and classes for sources in the
  reliable YSO catalog.  The South Pillars, West region, and total
  values are shown by dashed, dotted, and solid lines, respectively
  (as in Figures~\ref{fig:klf} and \ref{fig:36lf}).  The black lines
  show relative numbers in Stages 0/I, II, and III according to the
  results from our model fits of SEDs (note that 432 sources in the
  South Pillars and 58 in the West region are not plotted because they
  were labeled with ``ambiguous'' stages, between Stages I and II or
  between Stages II and III).  Gray lines use the standard color-color
  criteria for Class I, II, and photospheres from Allen et al.\
  (2004).}\label{fig:class}
\end{figure}
%%%%%%%%%%%%%%%%%%%%%%%%%%%%%%%%%%%%%%%%%%%%%%%%%%%%%%%%%%%%%%%%%%%%%%

Most of the ``ambiguous'' stage YSOs are probably in between Stage 0/I
and Stage II.  Figure~\ref{fig:class} shows a histogram of the number
of YSOs classified in each stage by the SED fitter (black lines) as
compared to a standard classification as Class I, II, or III
(photospheric) based on color-color selection (gray lines), following
the criteria of Allen et al.\ (2004).  The dotted lines show the West
region, the long dashes are for the South Pillars, and the solid lines
are for both regions.  While the numbers of embedded protostars (Stage
0/I and Class I) are similar in both methods, the color-color
selection (which has no ``ambiguous'' stages) has vastly more Class II
sources than Stage II sources from the SED fitter.  Most of the
``ambiguous'' sources from the SED fitter would therefore have been
classified as Class II protostars, although we cannot classify them
reliably as disk-dominated or envelope-dominated without data at
longer wavelengths.

Contamination of the YSO catalog by background dusty active galactic
nuclei (AGN) is a concern, because they can have similar colors.
Harvey et al.\ (2007) characterized the extragalactic contaminants in
{\it Spitzer} c2d images of the Serpens star forming region and found
that background star-forming galaxies and dusty AGN are generally
faint, with [4.5] $>$ 13.05 mag. In our YSO catalog, 87/907 ($<$10\%)
of sources fall below this [4.5] cutoff. Carina is $\sim$10 times more
distant than Serpens and lies near the Galactic midplane ($|b|$ $<$
1\arcdeg). We therefore expect many low-mass YSOs in Carina to be
fainter than this cutoff. We also expect that the additional
extinction toward extragalactic sources renders them fainter in our
sample compared to the c2d sample. For these reasons, 10\% is a high
upper limit to the true extragalactic contamination fraction in our
catalog. Applying less conservative IRAC color-mag selection criteria
of Harvey et al.\ (2006) results in only 2 extragalactic
candidates. The true fraction of extragalactic sources in the YSO
catalog is likely to be well below a reasonable limit of $<$5\%.

%%%%%%%%%%%%%%%%%%%%%%%%% FIGURE 7 - KLf  %%%%%%%%%%%%%%%%%%
\begin{figure}\begin{center}
%\epsscale{0.99}
%\includegraphics[width=3.2in]{../Povich/klf.eps}
\includegraphics[width=3.2in]{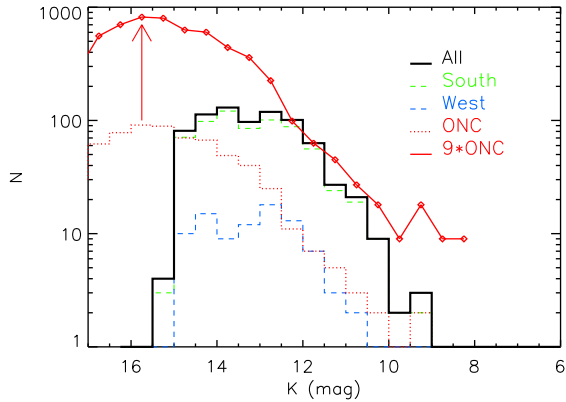}
\end{center}
\caption{$K$-band (2MASS) luminosity function (KLF) for the 953
  sources that were well fit by YSO SEDs by the fitting routine (cyan
  sources in Figure~\ref{fig:ysos}).  The black histogram is a sum of
  both the South Pillars and West region.  The red dashed histogram is
  the KLF for the Orion Nebula Cluster from Muench et al.\ (2002) with
  the apparent magnitudes shifted appropriately for a distance of 2.3
  kpc, while the solid red histogram is the Orion KLF multiplied by a
  factor of 9 normalized to the larger number of sources detected in
  Carina.}\label{fig:klf}
\end{figure}
%%%%%%%%%%%%%%%%%%%%%%%%%%%%%%%%%%%%%%%%%%%%%%%%%%%%%%%%%%%%%%%%%%%%%%

%%%%%%%%%%%%%%%%%%%%%%%%% FIGURE 8 - Band 1 Lf  %%%%%%%%%%%%%%%%%%
\begin{figure}\begin{center}
%\epsscale{0.99}
%\includegraphics[width=3.2in]{../Povich/3.6lf.eps}
\includegraphics[width=3.2in]{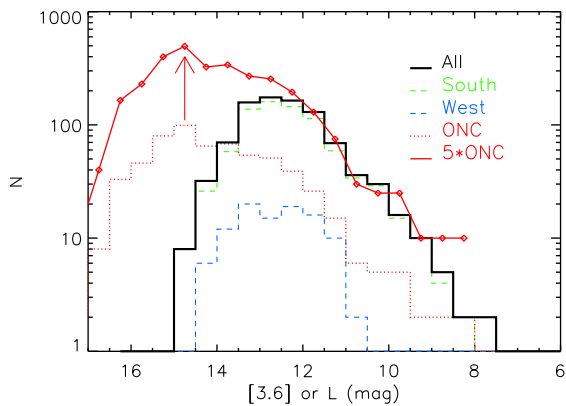}
\end{center}
\caption{Same as Figure~\ref{fig:klf}, but {\it Spitzer} Band-1
  (3.6~$\mu$m) luminosity function for the 909 sources that were well
  fit by YSO SEDs by the fitting routine (cyan sources in
  Figure~\ref{fig:ysos}).  Here the Orion $L$-band luminosity function
  (LLF; red) from Muench et al.\ (2002) was adjusted to the ppropriate
  distance and then multiplied by 5 to normalize to the number of
  sources Carina.}\label{fig:36lf}
\end{figure}
%%%%%%%%%%%%%%%%%%%%%%%%%%%%%%%%%%%%%%%%%%%%%%%%%%%%%%%%%%%%%%%%%%%%%%

\subsection{Spatial Distribution of YSOs}

It is immediately clear from Figure~\ref{fig:ysos} that the YSO
population in Carina is not evenly distributed across the area
observed by IRAC.  The vast majority of YSOs are concentrated in a
large cluster in the center of the image, occupying a 25\arcmin\
(17~pc) diameter cavity encircled by PAH emission from dust pillars.
These sources are a mix of Stage 0/I and Stage II, with many ambiguous
classifications that are confused between these two stages.  Most of
the sources are Stage II, however, suggesting ages of $\sim$1 Myr
(e.g., Haisch et al.\ 2001).  One gets the impression that 1 Myr ago,
this central cavity of the South Pillars was teeming with Class 0
protostars, molecular outflows, and dense cores.  Now the gas has been
mostly cleared from this central region, exposing the newly born
stars, except for the dense gas that remains in dust pillars.  The
next wave of star formation may occur in the molecular cloud
associated with the Giant Pillar (see Figure~\ref{fig:colorSP}) and
its extended features to the south.

Looking more closely at Figures~\ref{fig:ysos} and \ref{fig:stage},
one can also see significant sub-clusters of YSOs in the South Pillars
on size scales of $\la$2-6\arcmin\ (1-4 pc).  Many of these
sub-clusters of YSOs are associated with dust pillars, either embedded
within them or residing just outside the pillar heads.  These YSO
sub-clusters are also coincident with small clusters of point sources
(including those not identified as YSOs) seen in IRAC images, several
of which contain O-type stars.  The spatial relationship between these
small clusters and dust pillars will be discussed more in \S 4.5.  By
contast, the Western region has no large concentration of YSOs and
only a few very modest overdensities of YSOs on smaller scales.

Averaged across the entire South Pillars IRAC image, the YSO density
is 832 sources deg$^{-2}$, with much higher values in the central
concentration of YSOs noted above. In the West mosaic, by contrast,
the YSO density is 500 sources deg$^{-2}$ with little sign of
clustering.  The lowest-density region at the East edge of the South
Pillars has a YSO density of 265 sources deg$^{-2}$.  If this is
considered an upper limit to the background density of unassociated
background IR excess sources, then the vast majority of YSO sources
are indeed local to Carina.
%Even this ``background'' in the South Pillars is probably an
%overestimate with several of those sources being members of Carina,
%because it is much higher than the background density of 100 sources
%deg$^{-2}$ in a control field near M17, which is located in the inner
%Galaxy.  
We would expect the true background contaminant density in IRAC images
of the South Pillars and the West region to be $<$100 sources
deg$^{-2}$.

%------------------------------------------------------------------
\subsection{Summary of Results for Point Sources}

Of the more than 40,000 sources that we detect in 4 or more filters in
the merged 2MASS+IRAC point source catalog, less than 1,000 are highly
reliable YSOs based on fits to the SEDs.  The vast majority of sources
are foreground or (more numerous) fainter background sources.  Our
line-of-sight to Carina looks down the tangent point in the Carina
spiral arm, providing a high column density of stars for several kpc
behind Carina.  The known stellar population within Carina includes
about 70 O-type stars and (5--8)$\times$10$^4$ total stars (Smith
2006; Smith \& Brooks 2007), and the majority of these are
concentrated in the first-generation 2--3 Myr old clusters (Tr~14 and
16).  (Smith \& Brooks derived the total number of stars by
extrapolating to low masses using a Trapezium-like initial mass
function [IMF] from Meunch et al.\ 2002.)  Considering that the YSO
lifetime is a few 10$^{5}$ yr for the stars of 2--3 $M_{\odot}$ that
dominate our sample, one would expect of order several 10$^3$ YSOs in
Carina if star formation has been constant.  We detect almost 1000
YSOs, so it would appear that star formation is less vigorous now than
in the past.  However, the South Pillars and West region do not cover
the full nebula, and our SED-based selection of YSOs is highly
incomplete (see below), so perhaps the number of YSOs in our sample is
roughly consistent with continuous star formation averaged over time
(see \S 5).

An important point to emphasize is that the number of YSOs listed in
our catalog is certainly an underestimate of the true YSO population.
One reason for this is because of our rigorous criteria for
classifying the SEDs and because of complications with extended
emission.  Our sample was selected for reliability rather than
completeness; the SED-based selection is conservative, making us far
more likely to exclude true YSOs than to include false IR excess
sources.  Also, bright diffuse PAH emission from pillars can cause
spurious excess 8 $\mu$m emission, causing us to reject sources.
Furthermore, some of the more massive YSOs are in deeply embedded
clusters with bright diffuse emission, and as extended objects, they
were simply not included in our {\it point source} catalog to begin
with.  For example, IRAS 10430 is regarded as a luminous embedded
cluster and was the first sign of ongoing star formation recognized in
Carina (Megeath et al.\ 1996), but it is a bright extended source and
is not in our point source catalog, so the SED fitter identified zero
YSOs there.  Similarly, the SED fitter recognized zero YSOs in the
Treasure Chest cluster, mostly due to diffuse background emission.
However, detailed study of the $JHK$ photometry of this cluster
revealed an extremely high disk excess fraction of 67\%, with over 100
disk sources, a few moderately massive stars, and an age for the
cluster of $\la$10$^5$ yr (Smith et al.\ 2005).  Thus, embedded
clusters and massive YSOs with extended structure will be
systematically excluded from the YSO catalog.  Because of the
difficulty introduced by extended emission, many of these sources must
be dealt with on a case-by-case basis in future papers.  Lastly, even
without problems due to extended emission, our catalog is biased
against the youngest Class I and Class 0 protostars because those are
very red and are only likely to be detected at the longest
wavelengths, making it difficult to meet our selection criteria of
detection in at least 4 out of 7 filters.

More seriously, we are also missing many faint low-mass YSOs due to
sensitivity limits.  Figures~\ref{fig:klf} and \ref{fig:36lf} show the
$K$-band luminosity functions (LFs) and 3.6 $\mu$m LFs for the South
Pillars and the West region.  Both show a very steep power-law slope
with turnovers at $m_K \simeq$ 13 mag and $m_{3.6} \simeq$ 12.5 mag due
to completeness.  This is evident when the Carina LFs are compared to
scaled LFs from the Orion Nebula Cluster in Figures~\ref{fig:klf} and
\ref{fig:36lf} (from Muench et al.\ 2002).  Model SED fits suggest
that we are incomplete below stellar masses of 2--3 $M_{\odot}$.
Correcting for the missing low-mass sources by extrapolating a scaled
Trapezium-like IMF (Meunch et al.\ 2002; the same extrapolation used
by Smith \& Brooks 2007) would imply a total number of at least 5,000
YSOs in the South Pillars and West region alone, corroborating our
initial expectations above.  Given that this number of YSOs is roughly
10\% of the total number of stars in the Carina Nebula (Smith \&
Brooks), and that the YSO lifetime is roughly 10\% of the lifetime of
the region, we find it likely that star formation may have continued
at a relatively constant rate over the lifetime of the region.

Although we do not model the observed mass function directly from our
data, we can glean some clues about the IMF for ongoing star formation
in Carina from Figures~\ref{fig:klf} and \ref{fig:36lf}.  In the
magnitude range where our sample is fairly complete, the slopes of the
$K$ and 3.6 LFs in Carina are comparable to those in Orion, except at
the brightest end where there appears to be a slight excess of sources
in the scaled Orion sample.  Note, however, that the top few bins in
the Orion LFs are scaled from detections of only 1--2 sources each.
We therefore find no compelling evidence that the IMF of sources
formed recently in the South Pillars of Carina differs significantly
from the IMF of the Orion Nebula Cluster.  In addition, we must recall
that our Carina sample is based on YSOs selected by their SEDs,
whereas the Muench et al. sample refers to all stellar sources (stars
with and without disks).  If more massive stars shed their own disks
more quickly, this may account for the slight difference between the
bright ends of the Orion and Carina LFs.  Recall that we find 15--20
O-type stars among the South Pillars in Carina, none of which were
identified as YSOs from their SEDs.  Unfortuntely, the sensitivity
limits of our data do not permit us to comment on the lower end of the
IMF, which is potentially the most interesting (i.e., one would like
to know if the turnover mass of the IMF is skewed to higher masses in
externally-heated regions like this).

The number of embedded YSOs can also be compared to the number of
detected outflows in the Carina Nebula.  In an H$\alpha$ imaging
survey with the {\it Hubble Space Telescope} ({\it HST}), Smith et
al.\ (2010) detected 39 jets and jet candidates, 28 of which overlap
with the south and west regions covered by the IRAC images.  In the
IRAC data, we find roughly 160 YSOs that are Stage 0/I or Class I,
which is the YSO phase during which one expects to see ouflows.  The
28/160 ratio suggests that the lifetime for the most active
accretion/outflow phase during which an HH jet is detectable in a
region like Carina is 10--20\% of the duration of the embedded
protostar phase (Stage 0/I), or a few $\times$ 10$^4$ yr.  This is
commensurate with the observed dynamical timescale for some of the
longer HH jets (Smith et al.\ 2010, 2004a).  There are considerable
uncertainties, however, because both values are underestimates.  We
have noted that our YSO catalog misses many YSO sources below 2
$M_{\odot}$.  Also, Smith et al.\ (2010) found evidence that HH jets
with lower mass-loss rates are less easily detected and that the true
number of active HH jets in Carina (correcting also for the area not
surveyed by {\it HST}) may be of order 150--200.

%
%
%
%
%%
%%%%%%%%%%%%%%%%%%%%%%%%%%%%%%%%%%%%%%%%%%%%%%%%%%%%%%%%%%%%%%%%%%
%%%%%%%%%%%%%%%%%%%%%%%%%%%%%%%%%%%%%%%%%%%%%%%%%%%%%%%%%%%%%%%%%%
\section{Diffuse Emission}

\subsection{General Properties of the Extended Emission: Dust Pillars}

%As noted above for point sources, stellar photospheric emission
%contributes most to Band 1, while at longer wavelengths the point
%sources become increasingly dominated by disk emission for YSOs and by
%free-free wind emission for hot massive stars.  
In terms of spatially extended emission that contributes to the
various IRAC filters, Band 1 (3.6 $\mu$m) is dominated by diffuse
emission from the 3.3 $\mu$m PAH feature, Band 2 (4.5 $\mu$m) includes
the Br$\alpha$ line plus CO bandhead emission and H$_2$ lines, and
Bands 3 and 4 (5.8 and 8.0 $\mu$m) are dominated by strong PAH
features, although a few sources show extended thermal continuum
emission from warm circumstellar dust in these bands.

The most striking features in the IRAC images are the numerous dust
pillars and cometary clouds scattered throughout the region, most of
which harbor YSOs.  Their photodissociation regions (PDRs) that mark
the FUV-illuminated surfaces of molecular clouds are bright in Bands
1, 3, and 4, and so dust pillars appear magenta in typical IRAC color
images (Figures~\ref{fig:colorSP} and \ref{fig:colorWest}).  Similar
emission from dust pillars is widely seen in IRAC images of other
massive star-forming regions (e.g., Churchwell et al.\ 2004, 2009;
Reach et al. 2004, 2009; Bowler et al.\ 2009).

This emission from the South Pillars in Carina gives an unprecedented
view of the complex structures that arise when a clumpy molecular
cloud is shredded and destroyed by stellar winds and radiation from
massive stars, and acts as an excellent tracer of the overall
distribution of molecular clouds in the region.  The detailed
structure of the many South Pillars shows corrugations and long
twisted tails, most notably in the upper portions of the South Pillars
close to Tr~16.  The highly fragmented and filamentary structure of
the tails that trail behind dense knots indicates that Rayeigh-Taylor
and Kelvin-Helmholtz instabilities have efficiently disrupted their
parent clouds (e.g., Spitzer 1954; Williams et al.\ 2001).  These
structures are quite reminiscent of simulations of turbulent cloud
destruction by shocks (Pittard et al.\ 2009; Marcolini et al.\ 2005)
or instabilities at advancing ionization fronts (Whalen \& Norman
2008; Williams et al.\ 2001; Mizuta et al.\ 2006; Mackey \& Lim 2009).
Interestingly, the highly fragmented structures in Carina are more
closely matched by simulations that do not include efficient turbulent
mixing or efficient conduction, both of which tend to inhibit the
heavy fragmentation (Pittard et al.\ 2009; Marcolini et al.\ 2005).
These considerations are relevant to the survival times of these
features, and a more detailed comparison between observations of
pillars and simulations will be undertaken in a later paper.

Although the PAH emission traces the location of the cloud boundaries
very well, it arises mainly from a thin skin on the surface of these
clouds.  Kassis et al.\ (2006) estimated the thickness of the layer in
the Orion Bar to be of order 0.2 pc, corresponding to an unresolved
thickness of 2\arcsec\ in our Carina images.  The intensity of the PAH
emission does not necessarily trace the mass of molecular gas within
the clouds because much of the molecular gas is shielded from the FUV
radiation (Smith \& Brooks 2007; Brooks et al.\ 2000; Rathborne et
al.\ 2002).  The PAH emission also traces the warm dust in the PDRs,
matching maps of the 60 and 100 $\mu$m emitting optical depth (Smith
\& Brooks 2007), suggesting that IRAS emission comes mainly from thin
skins of these clouds as well, not far from the PAH emitting regions.
Sufficiently high spatial resolution images at longer wavelengths are
not yet available to verify the location of the far-IR-emitting dust.

The thermal continuum emission from warmer dust at 20--25 $\mu$m has a
different spatial distribution, located fully outside the dust pillars
and more closely matching that of the ionized gas seen in radio and
Br$\alpha$ emission (Smith \& Brooks 2007).  Some of the green glow
may also be due to hot dust intermixed with this ionized gas
(Churchwell et al.\ 2004; Povich et al.\ 2008).  Extended Br$\alpha$
is seen widely throughout the region, appearing green in color images
(Figures~\ref{fig:colorSP} and \ref{fig:colorWest}).  Br$\alpha$
emission arises mainly from diffuse gas in the open spaces between
dust pillars, rather than the ionization fronts on their surfaces.
Exceptional cases of extended Band 2 emission (green) and Band 3-4
emission (red) from individual sources are also seen, discussed below.

%%%%%%%%%%%%%%%%%%%%%%%%% FIGURE 9 - extended green emission  %%%%%%%%%%
\begin{figure*}\begin{center}
%\epsscale{0.99}
%\includegraphics[width=5.8in]{../BOTH_MOSAICS/subGreen.eps}
\includegraphics[width=5.8in]{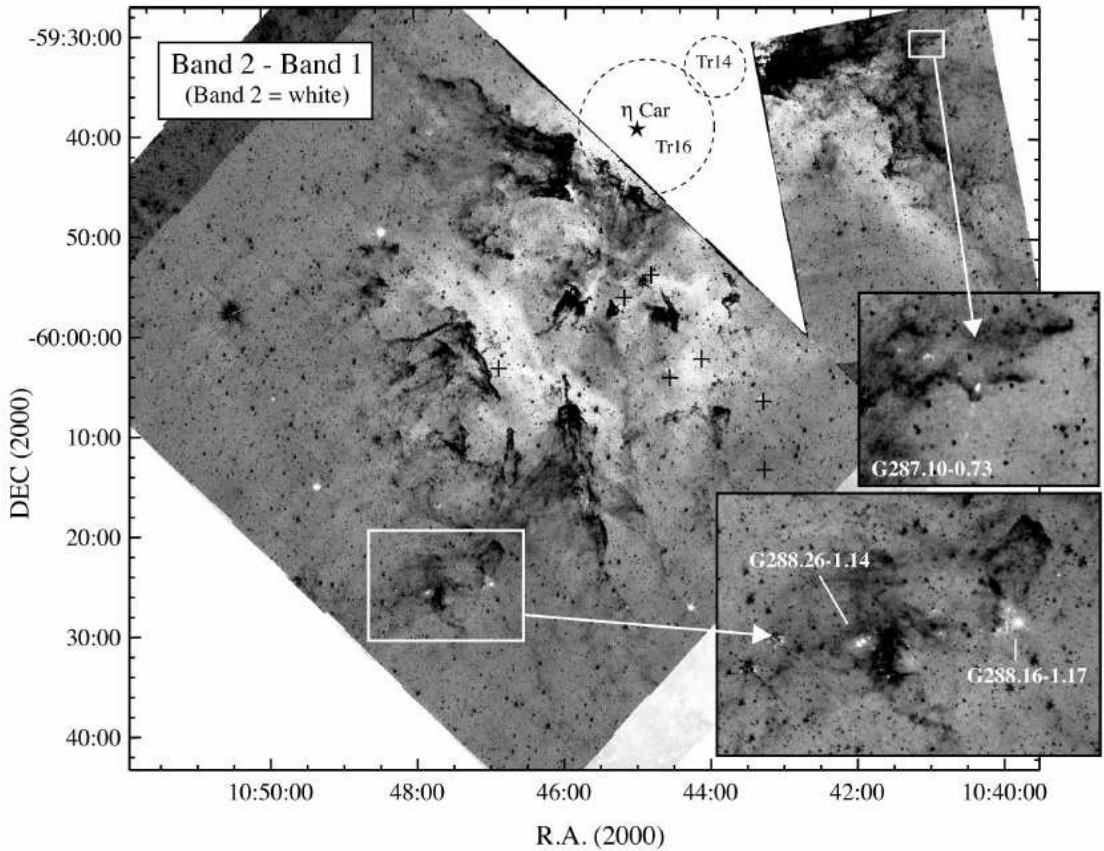}
\end{center}
\caption{A difference image made by subtracting the IRAC Band 1
  (3.6~$\mu$m) flux from the Band 2 (4.5~$\mu$m) image.  In the
  resulting difference image, extended emission that is black or dark
  gray represents excess 3.6~$\mu$m emission, dominated by the
  3.3~$\mu$m PAH emission feature included in the filter bandpass,
  accenting the dust pillars.  Most stellar point sources have a
  larger flux in the Band 1 filter and therefore appear black here as
  well; only highly reddened point sources or YSOs with strong excess
  dust emission will have a white color.  Diffuse regions appearing
  white or very light gray have excess emission in the Band 2 filter,
  due to either (1) Br$\alpha$ emission at 4.05 $\mu$m, or (2) H$_2$
  or CO bandhead emission from molecular outflows.  The two insets
  zoom in on regions with possible molecular outflows (see text).
  Crosses mark the positions of EROs (shocks), discussed in \S
  4.3.}\label{fig:green}
\end{figure*}
%%%%%%%%%%%%%%%%%%%%%%%%%%%%%%%%%%%%%%%%%%%%%%%%%%%%%%%%%%%%%%%%%%%%%%

\subsection{Extended Green Objects (EGOs) and Outflow Activity}

Previous studies of star-forming regions with IRAC have revealed a
population of extended objects that exhibit excess emission in the
4.5~$\mu$m Band 2 filter, as compared to adjacent filters.  These
features appear green in the common color coding of 3.6~$\mu$m (blue),
4.5~$\mu$m (green), and 8.0~$\mu$m (red) used to display IRAC images,
and so these have been referred to as ``extended green objects''
(EGOs).  This green emission is seen from low-mass Herbig-Haro (HH)
jets when the jet burrows into a molecular cloud, as in the famous
case of HH 46/47 (Noriega-Crespo et al.\ 2004; Velusamy et al.\ 2007)
and numerous jets in NGC~1333 (Gutermuth et al.\ 2008).  EGOs are also
closely asociated with outflows in regions of high-mass star formation
(Araya et al.\ 2007; Cyganowski et al.\ 2007; Davis et al.\ 2007;
Hunter et al.\ 2006; Shepherd et al. 2007; Smith et al.\ 2006).  The
emitting species is probably either CO bandhead emission or H$_2$
lines in the Band 2 filter, or both.  Cyganowski et al.\ (2008)
collected a large sample of EGOs seen in GLIMPSE data, and showed that
they tend to be correlated with infrared dark clouds (IRDCs) and with
CH$_3$OH maser emission, and that their associated continuum sources
have IRAC colors similar to the embedded accreting YSOs.  Cyganowski
et al.\ (2008) therefore suggested that most EGOs are good tracers of
molecular outflows specifically from high-mass protostars.

%-------------------------------
We detect few of the EGOs that are normally taken as signposts of
outflow activity in Spitzer data.  Figure~\ref{fig:green} shows a
subtraction image made from the difference between the Band 2 and Band
1 IRAC images.  Visual inspection reveals that the 4.5~$\mu$m excess
emission in this map (appearing white or light gray in
Figure~\ref{fig:green}) correlates well with extended features that
appear green in a composite color image.  Most of the diffuse green
emission that one sees in a color image is in the form of large areas
of diffuse emission, most likely due to Br$\alpha$ line emission from
photoionized gas in dense photoevaporative flows from the surfaces of
large dust pillars, or hot dust mixed with that gas, as noted above.
This appears as large white arcs draped over dark dust pillars in
Figure~\ref{fig:green}, with size scales of several arcminutes.  This
emission is well-correlated with large-scale H$\alpha$ and radio
continuum emission, as well as warm dust seen at $\ga$20~$\mu$m that
is heated {\it in situ} by trapped Ly$\alpha$ (see Smith \& Brooks
2007).  We detect only a few examples of the more compact diffuse
sources that one would associate with outflow activity.

There is no shortage of YSO outflow activity in Carina, however.  A
recent H$\alpha$ imaging survey with {\it HST} revealed 39 HH jets and
HH jet candidates in Carina (Smith et al.\ 2010).  Of the 28 HH
sources from Smith et al.\ (2010) that reside within the boundaries of
the South Pillar and western regions that we have surveyed with IRAC,
{\it zero} exhibit detectable excess 4.5~$\mu$m emission and so none
of them appear as EGOs.  This is somewhat surprising, since nearly all
the HH jets in Carina are seen to emerge from dust pillars
(i.e. plenty of associated molecular gas), and several of the HH jets
in this sample appear to mimic the case of HH~46/47, where one side of
a bipolar flow emerges into an H~{\sc ii} region cavity, while the
other burrows into the molecular cloud.  A possible explanation for
this missing green emission is that --- unlike the cases of HH 46/47
or the jets in NGC~1333 --- the HH jets in Carina are {\it irradiated}
HH jets, meaning that they are detectable primarily because the gas in
the jet is photoionized as it is bathed in UV light from dozens of
nearby O stars (Reipurth et al.\ 1998; Bally \& Reipurth 2001).  In
that case, the molecular gas that one normally would expect to see in
the coccoon surrounding an HH jet is probably dissociated by the same
UV radiation that ionizes the jet body.  There is also a potentially
severe observational limitation: The green emission that might
otherwise be seen from the portions of the jets that burrow into the
dense molecular clouds may be too faint to detect amid the extremely
bright PAH emission from the irradiated cloud surfaces.  Most EGOs
have been identified in IR dark clouds without bright PAH emission
(Cyganowski et al.\ 2007).  It therefore appears that in giant H~{\sc
  ii} regions such as Carina, where strong UV radiation fields
dominate, that EGOs do not provide good tracers of the YSO outflow
activity.  We speculate that the same feedback-dominated conditions
which lead to a shortage of EGOs may also partly explain the general
lack of ultra-compact H~{\sc ii} (UCHII) regions in Carina (Brooks et
al.\ 2001), since both require relatively long-lived, deeply embedded
environments for massive stars.  If massive stars form in the heads of
pillars and are quickly uncovered by the advancng ionization front,
these embedded conditions will not exist for long.  WE find it quite
unlikely that feedback inhibits the formation of high-mass stars
altogether, since the South Pillar region contains $\sim$20 O-type
stars, some of which are closely associated with YSOs and dust pillars
(see \S 4.4).

%%%%%%%%%%%%%%% TABLE x - EGO green fuzzies %%%%%%%%%%%%%%%%%%%%%%%%%%%%%%%
\begin{table*}\begin{minipage}{5.4in}
\caption{EGOs Identified in IRAC Images}\scriptsize
\begin{tabular}{@{}lcccccl}\hline\hline
EGO  &$\alpha_{2000}$ &$\delta_{2000}$ &SB$_2$ &IRDC? &YSO? &comment \\ \hline
%%-------------------------------------------------------------
G287.10-0.73 &10:41:14.3 &--59:32:39 &0.9--7.6  &no  &0 &collimated bipolar flow, optically dark pillar \\
G287.95-1.13 &10:45:42.0 &--60:17:31 &$\sim$29  &yes &1 &extended emission around point sources \\
G288.15-1.17 &10:47:00.6 &--60:25:35 &$\sim$1.5 &yes &3 &extended E of pt. source, IRAS 10451-6010 \\
G288.26-1.14 &10:47:51.9 &--60:26:22 &4--13     &yes &4 &several blobs, C$^{18}$O outflow (15) \\
\hline
\end{tabular}\label{tab:ego}
Note: $SB_2$ is the approximate range of excess surface brightness of
EGO features (MJy ster$^{-1}$) in the Band 2 (4.5~$\mu$m) filter, with
nearby background subtracted.  ``IRDC?'' refers to whether or not the
source appers as a dark cloud at 8.0 $\mu$m in the Band 4 IRAC image,
and ``YSO?'' refers to the number of unresolved YSO sources identified
by our automated fitting routine (see \S 3.1).
\end{minipage}
\end{table*}

%%%%%%%%%%%%%%%%%%%%%%%%% FIGURE 10 - extended green stamps %%%%%%%%%%
\begin{figure*}\begin{center}
%\epsscale{0.99}
%\includegraphics[width=5.2in]{../FIGS/egos.eps}
\includegraphics[width=5.2in]{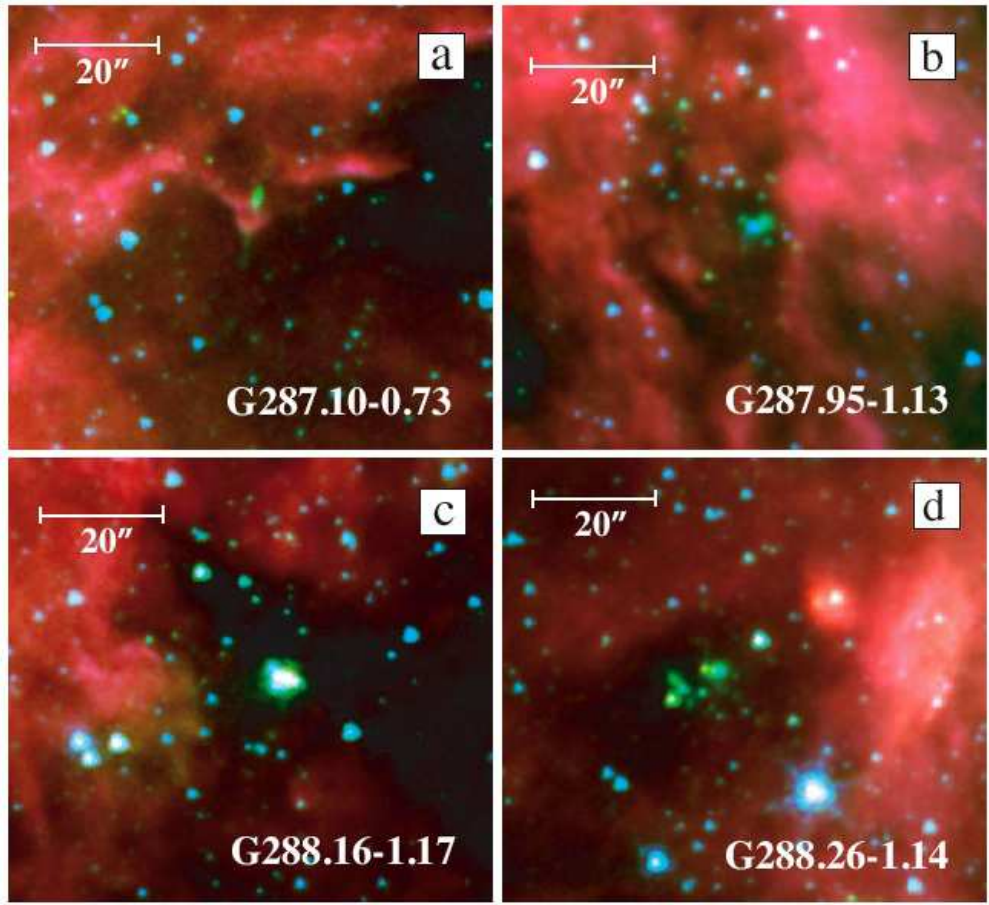}
\end{center}
\caption{Composite 3-color images of the four EGOs in Table 1, with
  IRAC Band1 (3.6~$\mu$m) in blue, Band 2 (4.5~$\mu$m) in green, and
  Band 4 (8.0~$\mu$m) in red.  Each panel has the same field of view
  with the size scale noted, oriented with north up and east to the
  left.}\label{fig:egos}
\end{figure*}
%%%%%%%%%%%%%%%%%%%%%%%%%%%%%%%%%%%%%%%%%%%%%%%%%%%%%%%%%%%%%%%%%%%%%%

There are nevertheless a few examples of EGOs in our IRAC survey of
Carina, none of which are associated with known HH jets.\footnote{EGOs
  seen in our IRAC survey were not targeted in the {\it HST} survey by
  Smith et al.\ (2010).  We could identify no sign of outflow activity
  in our ground-based optical images of the region, but many of the HH
  jets identified in {\it HST} images also could not be seen in
  ground-based images.  Therefore, it is possible that these outflows
  do in fact have associated HH objects that are undetected in
  ground-based wide-field images.}  These are shown in
Figure~\ref{fig:egos} and listed in Table~\ref{tab:ego}.\footnote{Many
  of the other compact features that appear white in
  Figure~\ref{fig:green} are not true EGOs, but artifacts from
  subtractions of bright point sources.  Point sources suffering large
  foreground extinction will have a relatively high 4.5/3.6~$\mu$m
  flux ratio in Figure~\ref{fig:green}.} The EGOs in Carina are found
preferentially in outer parts of the nebula away from the strongest UV
radiation, where their molecular outflows are presumably shielded from
dissociating radiation.  The four sources are described individually
below.

\smallskip

\noindent {\it G287.10--0.73}: This object, shown in
Figure~\ref{fig:egos}a, has the morphology of a collimated bipolar jet
oriented at P.A.$\simeq$350\arcdeg\ emerging from the end of a small
dust pillar.  The pillar is not an IRDC, but rather, has bright PAH
emission on its surface (Band 4) like most dust pillars.  In H$\alpha$
emission, the pillar is seen clearly in silhouette against the
brighter nebular background to the south, so it is not a highly
obscured background object.  The collimated jet body is seen as thin
green emission extending 5-10$\arcsec$ (0.05--0.1 pc) north and south
from the pillar head in color IRAC images, with the northern flow 7--8
times brighter in Band 2 emission.  The peak background-subtracted
surface brightness of the jet in Band 2 is 7.6 MJy ster$^{-1}$ at {\it
  Spitzer}'s spatial resolution, comparable to EGOs in GLIMPSE data
(Cyganowski et al.\ 2008).  The jet emission is not seen at visual
wavelengths (i.e. in H$\alpha$ or [S~{\sc ii}] images), although {\it
  HST} images of this position are not currently available;
nevertheless, the lack of bright H$\alpha$ in ground-based images (not
shown) suggests that this is not Br$\alpha$ emission in the IRAC Band
2 images, but rather, molecular emission from the outflow.
Presumably, the northern flow is brighter in Band 2 because in that
direction the jet burrows into the dust pillar, creating a coccoon of
molecular emission akin to that seen in HH~46/47 (Noriega-Crespo et
al.\ 2004).

\smallskip

\noindent {\it G287.95--1.13}: This EGO, shown in
Figure~\ref{fig:egos}b, is found amid a deeply embedded star cluster
on the west side of a giant dust pillar that harbors the most massive
molecular cloud in the South Pillars (Yonekura et al.\ 2005).  It
appears to be within an IRDC.  Some of the stars in this cluster
suffer high extinction and appear somewhat green (e.g., Indebetouw et
al.\ 2005), whereas others are blue in Figure~\ref{fig:egos}b, which
may imply that the dark cloud runs through the cluster.  Although the
central object here appears as extended green emission associated with
one YSO identified by our SED fitter (see \S 3.1), this may be a dense
cluster of highly embedded sources that is unresolved at the angular
resolution of {\it Spitzer}.  Ground-based 2-4 $\mu$m images with
higher spatial resolution could determine if this is truly an EGO or
unresolved stars.  If the green emission in Figure~\ref{fig:egos}b is
extended molecular emission in Band 2, it has a rather high peak
surface brightness of $\sim$20 MJy ster$^{-1}$.

\smallskip

\noindent {\it G288.16--1.17}: This EGO consists of extended green
emission within a few arcsec around a bright point source, shown in
Figure~\ref{fig:egos}c.  Careful inspection of the images, as well as
PSF subtraction, reveal that the Band 2 emission around the source is
definitely more spatially extended than in the other filters, mostly
to the east, and is not simply the PSF wings from a bright source.
The source is found within a very opaque IRDC that blocks most of the
surrounding Band 4 emission, and contains 3 YSOs identified by our SED
fitter.  The G288.16--1.17 EGO is coincident with IRAS 10451--6010,
which is a bright 25--60 $\mu$m source in deconvolved IRAS data, and
it resides at the western edge of the C$^{18}$O core number 15 in the
sample of Yonekura et al.\ (2005), although it is not coincident with
the outflow in that source.

\smallskip

\noindent {\it G288.26--1.14}: This EGO is a complex overlap of
several extended emission condensations within a $\sim$15\arcsec\
area, shown in Figure~\ref{fig:egos}d.  The green condensations have
background-subtracted peak surface brightness values of 10--13 MJy
ster$^{-1}$.  The most interesting aspect of this EGO is that it is
coincident with the bipolar molecular outflow seen in C$^{18}$O and
H$^{13}$CO+ emission (Yonekura et al.\ 2005), within the
2000~$M_{\odot}$ C$^{18}$O core number 15 that was identified as a
region of massive star formation.  The EGO is apparently found in an
IRDC, and no IRAS source is listed for this position, although there
is bright PAH and 25 $\mu$m continuum $\sim$30\arcsec\ to the west, at
the left side of Figure~\ref{fig:egos}d.  Our SED fitter identified 4
(+1 uncertain) YSOs in the central few arcsec amid the EGO emission,
indicating that this is a region of active ongoing star formation.
The coincidence with the CO outflow identified by Yonekura et al.\
(2005) gives high confidence that this EGO is indeed associated with
an outflow from ongoing {\it massive} star formation.

\smallskip

Finally, there were several sources that were revealed to have
extended excess 4.5/3.6 $\mu$m emission in Figure~\ref{fig:green}, but
which were not true EGOs because they had increasing extended excess
emission in Bands 3 and 4 as well.  Examples are the ring nebula
around SBW1 (see Smith et al.\ 2007), which is a SN~1987A-like ring
nebula surrounding a B supergiant, as well as arcs or bow-shock
structures around HD~305536 and HD~93222, which have late-O spectral
types.  In these cases, the extended Band 2 emission is likely to be a
combination of circumstellar Br$\alpha$ emission and thermal dust
emission warmed by direct radiation from the massive stars.  These
sources are discussed next.

%%%%%%%%%%%%%%% TABLE x - EROs red fuzzies %%%%%%%%%%%%%%%%%%%%%%%%%%%%%%%
\begin{table*}\begin{minipage}{6.1in}
\caption{EROs Identified in IRAC Images}\scriptsize
\begin{tabular}{@{}lccccccccl}\hline\hline
ERO &$\alpha_{2000}$ &$\delta_{2000}$ &SB$_4$ &star &Sp. Type &SED Type &A$_V$ &P$_{II}$ &comment \\ \hline
%%-------------------------------------------------------------
S1  &10:43:09.3 &--60:17:22 &12.5 &?            &?      &B5.5 V &2 &... &no SIMBAD object within 40\arcsec \\
S2  &10:43:17.9 &--60:08:03 &26   &HD~93027     &O9.5 V      &O8.5 V &1.4 &1.8 &Cr 228    \\
S3  &10:43:20.3 &--60:13:01 &17.6 &?            &?           &B2 V   &1.2 &... &2MASS J10432028-6013014 \\
S4  &10:44:11.1 &--60:03:21 &183  &HD~305536    &O9 V        &O8 V   &2   &4.8 &sp./ binary, Cr 228\\
S5  &10:44:36.3 &--60:05:30 &290  &HD~93222     &O7 III((f)) &O7 III &2.3 &8.6 &Cr 228 \\
S6  &10:44:50.5 &--59:55:45 &304  &CPD~-59 2605 &B           &YSO    &0.4 &... &Cr 228  \\
S7  &10:45:13.5 &--59:57:53 &229  &HD~305533    &B0.5 V      &B0.5 V &2.7 &... &Cr 228  \\
S8  &10:46:53.8 &--60:04:41 &52   &HD~93576     &O9 IV       &O8 V   &2.5 &... &sp.\ binary, Bo 11-9 \\
(S9)&10:44:05.1 &--59:33:41 &...  &(in Tr 14)   &B           &...    &... &... &Ascenso et al. \\
\hline
\end{tabular}\label{tab:ero}

Note: $SB_4$ is the approximate representative surface brightness of
ERO features (MJy ster$^{-1}$) in the Band 4 (8.0~$\mu$m) filter, with
nearby background subtracted, and ``star'' is the name of the central
star if known. ``Sp.\ Type'' is the observed spctral type, whereas
``SED Type'' is the stellar classification returned by the SED fitter;
A$_V$ is the corresponding visual magnitude from the SED fit.
P$_{II}$ is the ambient pressure in the H~{\sc ii} region in 10$^{-9}$
dynes cm$^{-2}$ (see text).
\end{minipage}
\end{table*}

%%%%%%%%%%%%%%%%%%%%%%%%% FIGURE 11 - extended red stamps %%%%%%%%%%
\begin{figure*}\begin{center}
%\epsscale{0.99}
%\includegraphics[width=5.5in]{../FIGS/shocks.eps}
\includegraphics[width=5.5in]{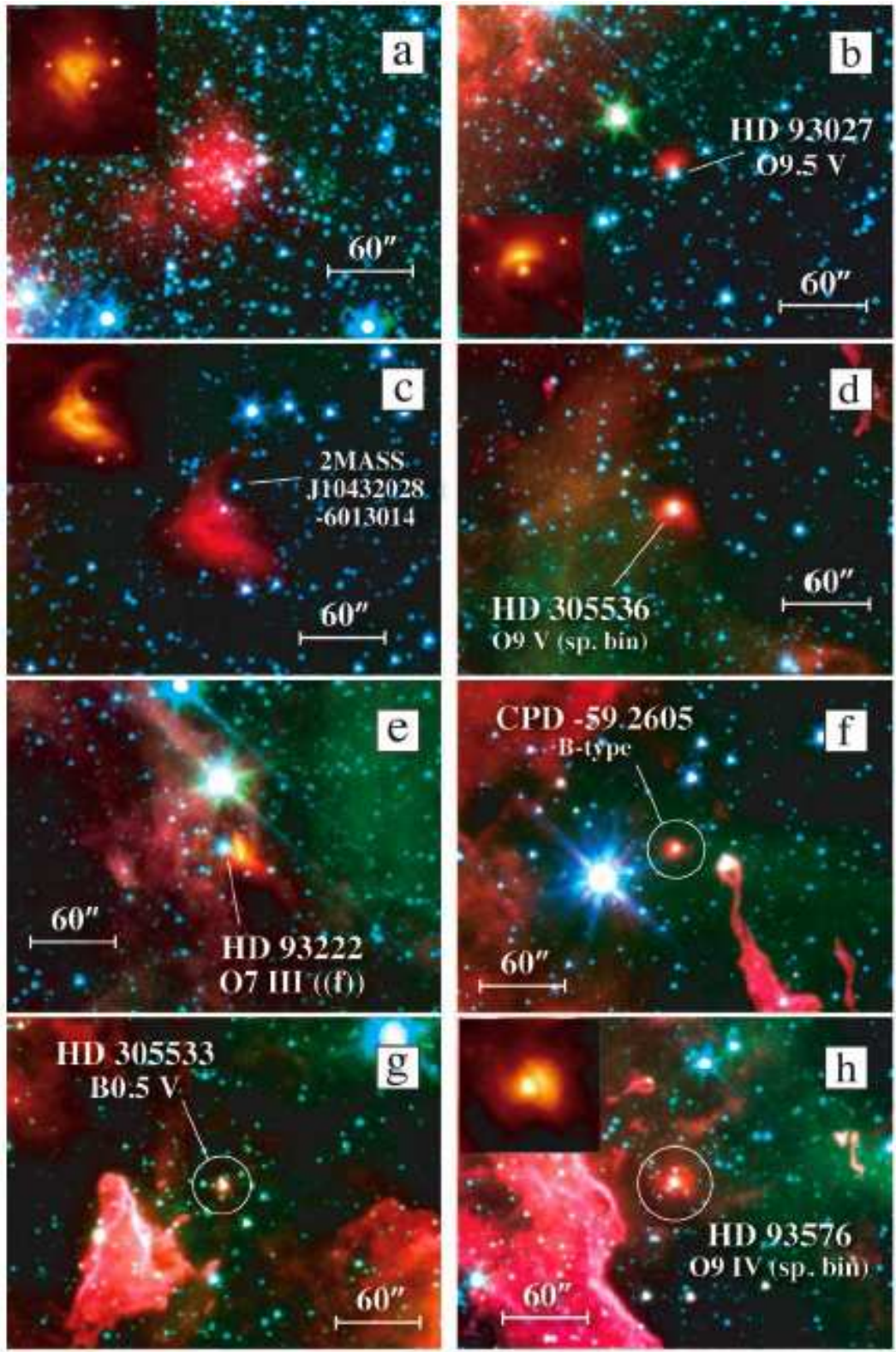}
\end{center}
\caption{Composite 3-color images of the eight EROs in Table 3, with
  IRAC Band 1 (3.6~$\mu$m) in blue, Band 2 (4.5~$\mu$m) in green, and
  Band 4 (8.0~$\mu$m) in red.  Each panel has the same field of view
  with the size scale noted, oriented with north up and east to the
  left.  The insets in Panels a, b, c, and h show false-color images
  in the Band 4 filter alone to show some of the faint shock structure
  that is not visible in the 3-color images.}\label{fig:shocks}
\end{figure*}
%%%%%%%%%%%%%%%%%%%%%%%%%%%%%%%%%%%%%%%%%%%%%%%%%%%%%%%%%%%%%%%%%%%%%%

\subsection{Extended Red Objects: Dusty Shocks}

Distinct from EGOs, there is also a clear population of extended {\it
  red} objects (ERO) that appear very red in the same 3-color IRAC
images. The deep red or reddish-orange color of these objects
(indicating increasing excess in Bands 2, 3, and 4) is distinct from
the magenta color that arises from the more widely seen PDR emission
at the surfaces of all molecular clouds in these same images, which
have bright PAH emission features contributing to the flux in Bands 1
and 4 (PAH features at 3.3, 7.7, and 8.6 $\mu$m).
Figure~\ref{fig:shocks} shows a sample of eight EROs identified in
IRAC images, also listed in Table~\ref{tab:ero}.  (These are the
clearest cases; there are other candidates which are more dubious
examples because of spatial resolution criteria rather than color
criteria.)  Inspection of the Band 4 images clarifies that these are
not simply the result of a broader PSF at 8.0~$\mu$m, because the
objects are clearly spatially extended far beyond the PSF shape.  The
objects extend $\sim$10\arcsec\ to 40\arcsec\ from a bright central
star seen at shorter wavelengths, and they are often one-sided and
shaped like bow-shocks, usually pointing to the interior of the H~{\sc
  ii} region.\footnote{Since the arcs point to the interior of the
  H~{\sc ii} region, we consider it unlikely that these are bow shocks
  around runaway stars.  In that case all sources would need to have
  been ejected from the periphery of the region and would be
  converging on the center.}

For the two objects in Panels (a) and (c) of Figure~\ref{fig:shocks},
the central star is relatively faint and not identified or the
spectral type is unknown.  These two do, however, show morphology that
is suggestive of curved stand-off shocks, especially in the 8.0~$\mu$m
images in the insets of these panels. Fits to their SEDs suggest
stellar types of B5.5 V and B2 V.

It is quite interesting that all other objects in
Figure~\ref{fig:shocks} (Panels b, d, e, f, g, and h) show a
consistent trend that the ERO nebulosity surrounds a moderately
massive, bright star known to be an early B-type or late O-type star
in the Cr~228 or Bo 11 clusters (see the list in Smith 2006, and
references therein). Four of the six are O-type stars, and two of
those have especially strong winds with spectral types of O7~III((f))
and O9~IV.  The stellar types inferred from SED fits agree remarkably
well with the observed spectral types for these stars (Table 3),
adding reliability to the B types inferred from the SEDs of S1 and S3,
as noted above.\footnote{Note, however, that the SED fitter identified
  the B star CPD -59~2605 as a YSO with low interstellar reddening,
  rather than a B star with reddening comparable to the other
  sources.} This is undoubtedly an important clue to the physical
origin of these shock structures, and suggests that they are similar
to the bow shocks around O stars seen in IRAC images of M17 and RCW~49
(Povich et al.\ 2008).  The fact that these stars can have substantial
stellar winds has obvious bearing on the clear bow-shock morphology
seen in several of the EROs in Figure~\ref{fig:shocks}.  The EROs S2
and S4 around HD~93027 and HD~305536 (both late O-type stars) are very
clear one-sided bow shocks with their apex pointing up toward Tr~16.
The EROs S6--S8 around CPD~-59~2605 and HD~305533 (both early B-type)
and the O9~IV star HD~93576 (located within the cluster Bo 11;
Fitzgerald \& Mehta 1987) are less extended, more fuzzy, and only
somewhat asymmetric.  These could be bow shocks viewed at a low
inclination (i.e. face-on).  The ERO S5 around the O7~III((f)) star
HD~93222 is more complex and extended, with a bright filament
extending to the south-west, but with a series of curved filaments in
other directions as well, forming an apparent fragmented shell around
the star.  This star is noteworthy because it has some of the highest
velocities observed in interstellar absorption lines among the O-type
stars in Carina (Walborn et al.\ 2002b, 2007; and references therein),
and we conjecture that these absorption features probably arise in the
dense post-shock gas associated with the ERO emission.

We suggest that most of these features arise from dusty shock fronts
that occur at the stand-off shock between the fast outflowing stellar
wind of the central star and the large-scale flow of plasma in the
surrounding H~{\sc ii} region.  The bulk flow of ambient plasma is the
result of photo-evaporative flows from the surfaces of nearby
molecular clouds, ablated and accelerated outward by feedback from
massive stars in Carina.  (S1, however, may be slightly different;
perhaps a B star embedded in the PDR, since the SED of its diffuse red
emission is different from the other sources.)  Such flows are known
to be present based on other tracers such as the many bent Herbig-Haro
(HH) jets and LL~Ori objects that serve as ``wind-socks'' in the
region (Smith et al.\ 2010), as well as pervasive diffuse X-ray
emission (Townsley et al.\ 2006).

The origin of the dust in these ERO shocks is uncertain, since the
fast and hot winds of O and early B stars are not efficient dust
producers.  The dust may simply pile-up at the dense shock front after
having been entrained by the photoevaporative flow off the surface of
a nearby molecular cloud.  Alternatively (but perhaps less likely
because of the very high required densities), the dust may actually
form in the post-shock cooling zone as is seen in some colliding-wind
binaries (e.g., Williams et al.\ 1990; Usov 1991; Crowther 2003; Smith
2009) and supernovae (Smith et al.\ 2008).  To be bright in IRAC Band
4, the dust must be heated to roughly 200--300 K.  The dust could be
heated by direct stellar radiation from the OB stars, it could be due
to trapped Ly$\alpha$ heating if it is mixed with dense ionized gas,
or the dust may be heated by collisions in the hot post-shock
gas.\footnote{Everett \& Churchwell (2010) infer very efficient dust
  cooling under similar conditions, making it difficult to maintain
  such high dust temperatures from radiation alone.  This may argue
  for a contribution of shock heating in these sources.}  EROs do
appear to be mixed with dense ionized gas, since 6 of the 8 sources
are projected amid locations with bright diffuse Br$\alpha$ emission
(see Figure~\ref{fig:egos}).  It would be interesting to investigate
possible correlations with diffuse X-ray emission in future studies.

Two of the objects (HD~305536 and HD~93576) are known spectroscopic
binaries.  Although WC+O binaries are famously known to produce dust
in their colliding-wind shocks, these sources are not WC binaries.
Furthermore, it does not seem likely that binarity is an important or
necessary condition for the circumstellar dust in an ERO, since other
objects in Table~\ref{tab:ero} are not known to be binaries.  The two
close binaries out of six early OB stars are consistent with a random
sample.

%%%%%%%%%%%%%%%%%%%%%%%%% FIGURE 12 - arrows %%%%%%%%%%
\begin{figure*}\begin{center}
%\epsscale{0.99}
%\includegraphics[width=5.5in]{../BOTH_MOSAICS/pillars.eps}
\includegraphics[width=5.5in]{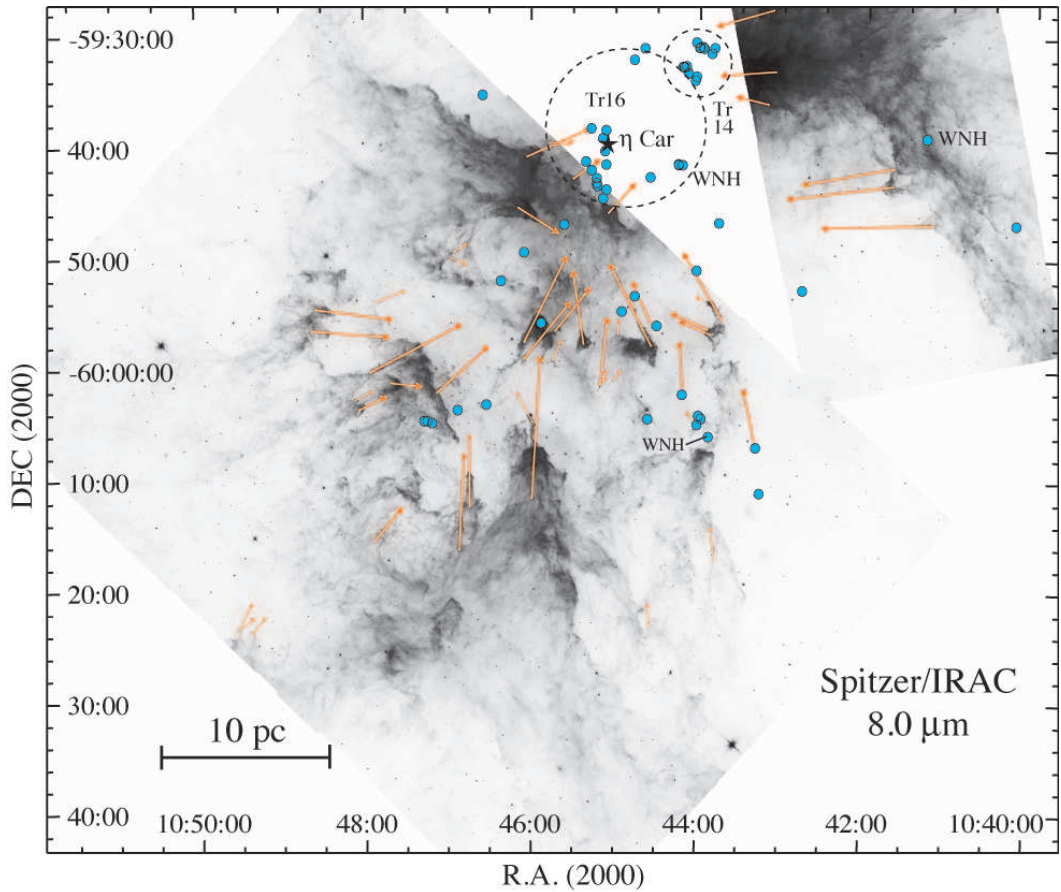}
\end{center}
\caption{The combined Band 4 (8 $\mu$m) IRAC image of the South
  Pillars and the western region, showing the structure of dust
  pillars seen in PAH emission.  The locations of O-type stars and WNH
  stars compiled by Smith (2006) are marked with blue circles.  The
  orange arrows mark the direction of the pillar axis for several
  well-defined dust pillars (in some cases small arrows mark
  directions for smaller pillars that may have a different direction
  from their parent pillar).  Notice that not all arrows point to
  $\eta$ Car and the Tr 14/16 clusters, located at the top of the
  image.}\label{fig:pillars}
\end{figure*}
%%%%%%%%%%%%%%%%%%%%%%%%%%%%%%%%%%%%%%%%%%%%%%%%%%%%%%%%%%%%%%%%%%%%%%

Given the large number of O and early B-type stars in Carina (see
Smith 2006), it might seem surprising that only six clear examples of
such dusty wind collision shocks are detected in the region we
surveyed with IRAC.\footnote{The IRAC images do show many examples of
  bow-shaped structures around bright point sources, but many of these
  are consistent with limb-brightened PAH emission from the heads of
  dust pillars that contain a YSO, where the Band 4 emission (along
  with emission in Bands 1 and 3) arises in a PDR at the head of the
  pillar.  These sources are not believed to be dusty wind collision
  shocks.} For example, there are no such structures surrounding
HD~92740 or HD~93131 (e.g., Crowther et al.\ 1995), which are extreme
WNH stars (Smith \& Conti 2008) with powerful stellar winds that are
located within the region we surveyed with IRAC.  Thus, these EROs in
Figure~\ref{fig:shocks} may arise from a combination of special
conditions: The central O-type or early B-type star with a strong wind
must be located in the outer periphery of the nebula near the surfaces
of molecular clouds, whereas many OB stars reside in the central
clusters Tr~14 and Tr~16 where much of the cooler gas has already been
cleared away; relatively few reside in the South Pillars.  Also, the
star must not be so luminous that it vaporizes the circumstellar dust,
and its wind must not be so strong that all material is cleared away
beyond distances where it can be heated.  Perhaps these reasons favor
relatively {\it late} O-type stars rather than the most massive WNH
stars or early O-type stars.  In addition, the star must be embedded
within a particularly dense photoevaporative flow.

It is obvious from inspection of the images that local density
enhancements are strong in the gas surrounding the South Pillars, and
several of the EROs do seem to be embedded in dense ionized gas as
noted above. In the three cases where a bow shock or arc is separated
from the star and spatially resolved, and where the star's spectral
type is known (i.e., EROs S2, S4, and S5) we can use the stellar wind
parameters and observed bow shock properties to infer the pressure in
the local ISM.  The pressure of the photoevaporative H~{\sc ii} region
flow, balanced with the stellar wind, is given by

\begin{equation}
P_{II} = \frac{\dot{M}_w v_w}{4 \pi R_s^2} ,
\end{equation}

\noindent where $\dot{M}_w$ and $v_w$ are the mass-loss rate and speed
of the stellar wind, and $R_s$ is the observed stand-off radius of the
bow shock.  Taking $\dot{M}_w$ and $v_w$ values for HD~93027,
HD~305536, and HD~93222 from the tables in Smith (2006; these
$\dot{M}_w$ values account for wind clumping), and observed radii of
the shocks from the star, we list corresponding values for $P_{II}$
for these three sources in Table~\ref{tab:ero}. For $R_s$, we took the
difference in radius between the star and the peak of the Band 3 and
Band 4 emission.  This is probably the largest source of uncertainty
(roughly $\pm$30\% in $P_{II}$), as the relative position of the shock
front and the dust is unclear.  In any case, the pressures derived for
the ambient medium are of order a few $\times$ 10$^{-9}$ dynes
cm$^{-2}$, which is commensurate with the pressure inferred from
ionization fronts in the H~{\sc ii} region by Smith et al.\ (2004b).

One might expect that the EROs in Figure~\ref{fig:shocks} are the most
extreme cases, and that there could be many more examples of such
stand-off shocks that are fainter or closer to their stars. {\it HST}
images of Carina reveal several very small ($\sim$0.05 pc) bow-shock
structures surrounding fainter stars that are embedded within dense
photoevaporative flows from dust pillars (Smith et al.\ 2010).

%%%%%%%%%

In Table~\ref{tab:ero} we list one additional object that is also a
likely member of the class (S9), but which was not included in the
field of view of our IRAC images.  Ascenso et al.\ (2007) discussed a
bow-shaped structure around a B-type star in the Tr~14 cluster in
near-IR $JHK$ images, and they noted that the apex of the bow points
toward the massive O2 star HD~93129A.  Ascenso et al.\ (2007)
considered it likely that this was a compact H~{\sc ii} region
associated with this star.  However, its size of $\sim$10\arcsec\ is
comparable to the other IR shocks in Table 3, and it is seen in IR
continuum images but is {\it not} detected in H$\alpha$ images
obtained with {\it HST} (Smith et al.\ 2010).  This argues that it may
be a dusty bow shock like the other objects in Table 3, and its
early-type star is consistent with other members of the group.

%%%%%%%%%%%%%%%%%%%%%%%%% FIGURE 13 - clusters %%%%%%%%%%
\begin{figure*}\begin{center}
%\epsscale{0.99}
%\includegraphics[width=5.5in]{../BOTH_MOSAICS/clusters.eps}
\includegraphics[width=5.5in]{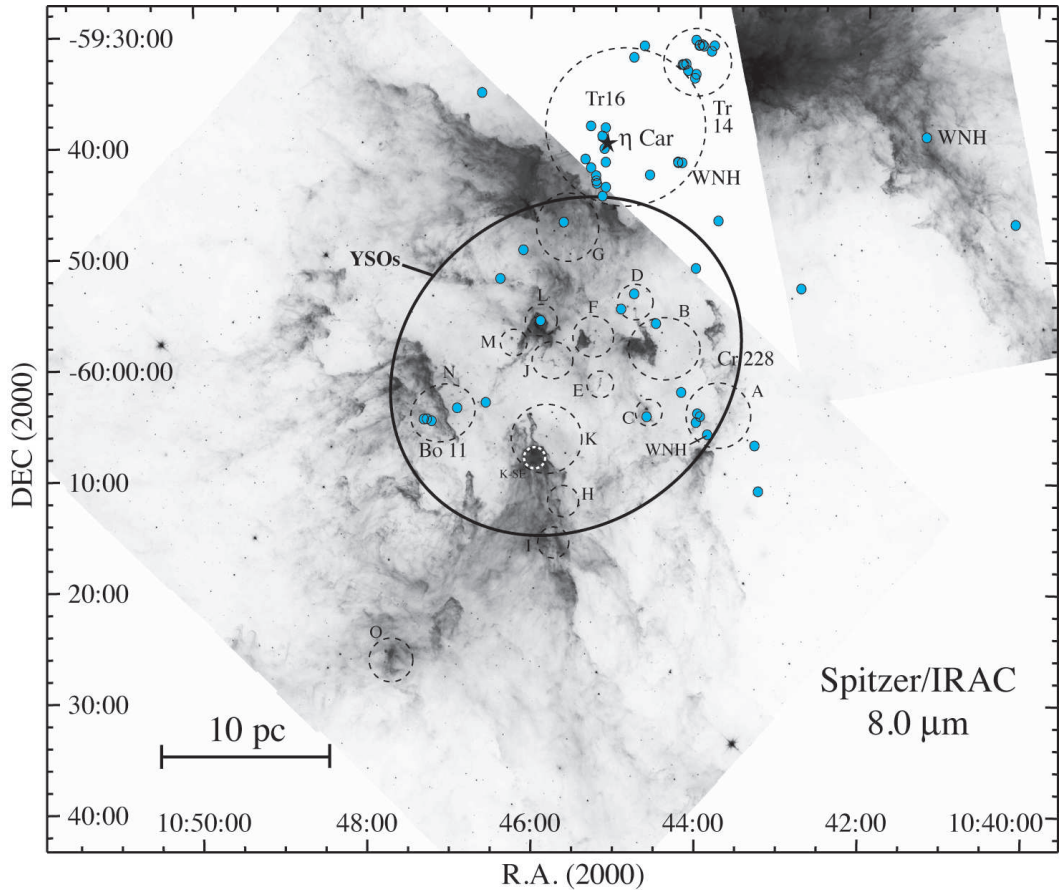}
\end{center}
\caption{Same as Figure~\ref{fig:pillars}, but with thin dashed
  circles marking locations of over-densities of stars seen in the
  Band 1 and 2 images.  The large solid ellipse marks the area with
  the greatest concentration of YSO candidates in the IRAC images
  identified by the automated SED fitter routine (see
  Figure~\ref{fig:ysos}).}\label{fig:clusters}
\end{figure*}
%%%%%%%%%%%%%%%%%%%%%%%%%%%%%%%%%%%%%%%%%%%%%%%%%%%%%%%%%%%%%%%%%%%%%%
%%%%%
%%%%%
%%%%%%%%%%%%%%% TABLE x - clustes %%%%%%%%%%%%%%%%%%%%%%%%%%%%%%%
\begin{table*}\begin{minipage}{5.2in}
\caption{Clusters in the South Pillars in 3.6 $\mu$m IRAC Images}\scriptsize
\begin{tabular}{@{}lccccl}\hline\hline
Cl.  &$\alpha_{2000}$ &$\delta_{2000}$ &O stars &YSOs &comment \\ \hline
%%-------------------------------------------------------------
A   &10:43:41 &--60:03:50 &6 &11   &Cr 228, WNH, ERO-S2, S4 \\
B   &10:44:22 &--59:57:50 &1 &24   &Cr 228     \\
C   &10:44:33 &--60:03:30 &1 &8    &ERO-S5     \\
D   &10:44:41 &--59:53:50 &2 &10   &ERO-S6     \\
E   &10:45:09 &--60:01:00 &0 &16   &...        \\
F   &10:45:12 &--59:56:40 &0 &30   &...        \\
G   &10:45:33 &--59:47:00 &1 &30   &Tr 16-SE, X-ray cluster \\
H   &10:45:36 &--60:11:30 &0 &14   &...        \\
I   &10:45:42 &--60:15:20 &0 &17   &EGO 287.95-1.13 \\
J   &10:45:45 &--59:58:50 &0 &8    &...        \\
K   &10:45:48 &--60:06:00 &0 &35   &HH 903     \\
K-SE&10:45:55 &--60:08:00 &0 &6    &embedded   \\
L   &10:45:54 &--59:56:50 &1 &(104)&embedded, Treasure Chest \\
M   &10:46:12 &--59:57:10 &0 &7    &...         \\
N   &10:47:05 &--60:03:50 &5 &23   &Bochum 11   \\
O   &10:47:41 &--60:26:00 &0 &11   &embedded, EGO 288.26-1.14 \\
\hline
\end{tabular}\label{tab:cluster}
%Note: .......
\end{minipage}
\end{table*}
%%%%%%%%%%%%%%%%%%%%%%%%%%%%%%%%%%%%%%%%%%%%%%%%%%%%%%%%%%%%%%%%%%%%%%%%%

%%%%%
%%%%%
\subsection{Orientations of Dust Pillars and Massive Stars}

UV radiation drives the ionization fronts that evaporate clouds, and
strong stellar winds can ablate those photoevaporative flows or
destroy clouds directly by shocks.  While the hydrodynamics of this
interaction can be complex (Pittard et al.\ 2009; Klein et al.\ 1994;
Whalen \& Norman 2008), one generally expects dust pillars and the
heads of cometary clouds to point toward the OB stars that have shaped
them.  These structures therefore identify the most influential
massive stars, locally or globally (see Smith \& Brooks 2007).

Figure~\ref{fig:pillars} shows the 8.0 $\mu$m Band 4 IRAC image of the
South Pillars and western region in Carina.  The structures in this
image are dominated by PAH emission from the surfaces of the molecular
gas in dust pillars and cometary clouds.  Superposed on this image, we
also draw arrows (orange) for well-defined elongation axes of dust
pillars, and we plot the locations of known O-type stars in Carina
with filled blue circles (see the list in Smith 2006).  The O-type
stars in the South Pillars region are dominated by members of the
Cr~228 and Bo-11 clusters, but the O star population is spatially
dispersed.

Although the global ionizing flux and stellar wind momentum flux is
dominated by the stars in Tr~14 and Tr~16 (Smith 2006), it is clear
from Figure~\ref{fig:pillars} that not all pillars point exactly
toward $\eta$ Car and its immediate neighborhood of Tr~16.  While
instabilities inherent to the cloud destruction process can cause some
irregular structure, this cannot explain all the observed deviations.
In some cases pairs of adjacent pillars which are still in-tact point
in the same direction, and that direction is more than 30\arcdeg\ from
the core of the nebula.  Figure~\ref{fig:pillars} shows that there are
$\sim$20 O-type stars scattered amid the South Pillars.  These tend to
be less extreme, late O-type (as well as several early B-type stars;
not shown), and some of these appear to have sculpted their local
environs more than the more massive O-type stars in the central
clusters Tr~14 and 16.  In some cases, Tr~14 and 16 are evidently too
far away to compete with the local influence of the less massive O
stars that are nearby.

It is interesting that the two WNH stars located far outside Tr~16
(HD~92740 far to the west and HD~93131 in Cr 228 to the south) do not
appear to have had much influence on the shaping of dust pillars in
their vicinity.  There are no pillars that clearly point to them.
Since these stars have some of the strongest winds and highest
luminosities of any massive stars in Carina, their lack of influence
suggests that they may be projected in the foreground or background
compared to the surrounding molecular gas, or that they have only
recently arrived in their present locations.  Walborn (1995) notes
that they could have drifted to their current locations from the core
of Tr~16 in $\sim$3 Myr if they are moving at only 10 km~s$^{-1}$.

Close inspection of the sub-structure in several of the South Pillars
reveals an interesting phenomenon.  Some pillars or pillar complexes
will follow an elongation axis at large scales, but on smaller scales
one can find protrusions or cometary clouds that point in an
altogether different direction. Some of these are marked in
Figure~\ref{fig:pillars} with smaller and thinner arrows.  Examples
are the pillars associated with the jets HH~666, HH~903, and the
cluster of HH jet candidates HHc-4 to 8 (located within cluster ``F''
in Figure~\ref{fig:clusters}; see Smith et al.\ 2010); in general
these point toward $\eta$~Carinae, but they also show smaller features
pointing off-axis by more than 30\arcdeg.  This would imply the rather
strange set of circumstances that the smaller sub-structure is shaped
by a different UV or wind source than the larger pillar structure that
it is a part of.

The dominant source of UV radiation or winds that shaped a given
pillar may have changed during the lifetime of the pillar in order to
cause this.  This could have happened for a variety of reasons:

1. A new O-type star may have appeared in the immediate vicinity of
the pillar.  This may occur because the O star was just recently
formed or just recently emerged from its natal cloud, so that its UV
radiation and winds have just turned on.

2.  It may have appeared in the vicinity because it formed elsewhere
and drifted away from its birth site.

3. While $\eta$ Carinae was on the main sequence, it probably
dominated the UV luminosity and winds in the region (Smith 2006).
Late O-type stars or early B-type stars are distributed throughtout
the region, and these had relativly little influence compared to
$\eta$ Car.  However, on local scales they became more influential
when $\eta$~Car entered its post-main-sequence luminous blue variable
(LBV) phase and its ionizing UV radiation effectively shut off (this
occured because the star became cooler and because its present-day
stellar wind is so dense that it absorbs all of the star's ionizing
radiation; the dusty Homunculus Nebula ejected in 1843 absorbs all
remaining FUV radiation).  The drop in UV flux at the onset of $\eta$
Car's post-main sequence phase alone probably caused the total
ionizing UV radiation from Carina to drop by $\sim$25\% (Smith 2006).

Though the last mechanism is known to have occurred, it is difficult
to rule out the first two options for any particular source.  In any
case, the observed structures illustrate that the South Pillars are in
a dynamic environment, and both scenarios would require rather long
lifetimes of $\sim$1 Myr for the pillars.  The pillar lifetimes of
$\sim$1 Myr agree with the approximate ages of the Stage~II sources
left in their wakes.

\subsection{Star Clusters and Dust Pillars}

Since dust pillars are thought to be prime sites for ongoing or
triggered star formation, we investigate the relative spatial
distribution of dust pillars as compared to YSOs and young star
clusters or groups.  Figure~\ref{fig:clusters} compares the Band-4
(8.0 $\mu$m) IRAC image of the South Pillars and West region,
dominated by PAH emission on the surfaces of dust pillars, to the
locations of O-type stars (blue dots) and overdensities of stars.  The
dashed circles in Figure~\ref{fig:clusters} identify the locations of
small clusters or groups of stars seen in the Band-1 (3.6 $\mu$m) IRAC
image, in which the point source emission is dominated by stellar
photospheres.  The identification procedure for these clusters is
subjective, so we have only considered clear cases where there is a
high over-density of stars compared to the surrounding field.  The
coordinates of these clusters (corresponding to the centers of the
dashed circles, labeled A, B, C...) are listed in
Table~\ref{tab:cluster}, along with their numbers of O stars and YSOs,
cross identifications, and other information.  There are several
smaller overdensities of stars and YSOs as well, which were not as
clear as those listed in Table~\ref{tab:cluster}.  For example, there
is a small cluster of YSOs at the head of the dust pillar from which
the jets HH~1004 and 1005 emerge (Smith et al.\ 2010), located to the
east of the giant pillar, but it does not show a clear overdensity of
point sources in the 3.6~$\mu$m image.

YSOs are not identified explicitly in Figure~\ref{fig:clusters}, but
they can be seen in Figures~\ref{fig:ysos} and \ref{fig:stage}, and
compared to the locations of clusters A--O.  This is discussed further
below.  In Figure~\ref{fig:clusters} we show a large solid ellipse
noting the approximate region with the highest concentration of YSOs
(see Figure~\ref{fig:ysos}), most of which are Stage-II/Class-II.
Several qualitative conclusions can be surmized from
Figure~\ref{fig:clusters}:

%%%%%%%%%%%%%%%%%%%%%%%%% FIGURE 14 - Tr16 SE  %%%%%%%%%%
\begin{figure*}\begin{center}
%\epsscale{0.99}
%\includegraphics[width=5.8in]{../FIGS/clusterIRAC.eps}
\includegraphics[width=5.8in]{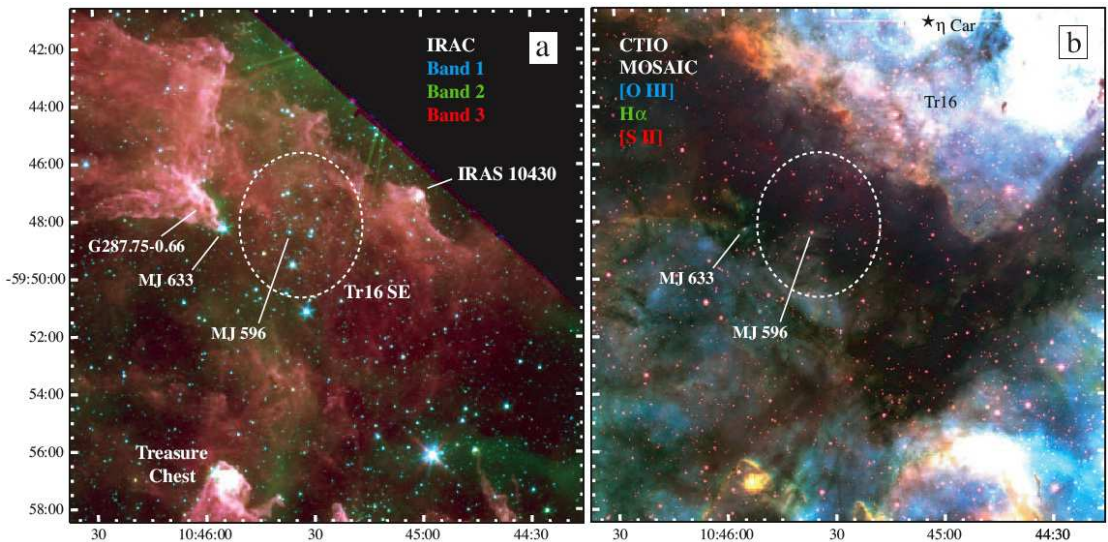}
\end{center}
\caption{Two images of the same field surrounding the embedded cluster
  Tr16 SE (cluster G in Figure~\ref{fig:clusters}).  (a) A 3-color
  composite image made with IRAC Bands 1 (blue), 2 (green), and 3
  (red).  (c) A color composite using optical narrowband images with
  [O~{\sc iii}] $\lambda$5007 \AA\ (blue), H$\alpha$ (green), and
  [S~{\sc ii}] $\lambda\lambda$6717,6731 (red).  The optical image was
  taken with the MOSAIC camera on the CTIO 4m telescope (see Smith et
  al.\ 2003).}\label{fig:Tr16se}
\end{figure*}
%%%%%%%%%%%%%%%%%%%%%%%%%%%%%%%%%%%%%%%%%%%%%%%%%%%%%%%%%%%%%%%%%%%%%%

1. Several of the sub-clusters in Figure~\ref{fig:clusters} seem to be
associated with both O-type stars and local concentrations of YSOs,
but there is a range of degree to this association.  For example,
cluster A (part of Cr~228) has 6 O-type stars (including one WNH) and
11 YSOs, cluster N (Bochum 11) has 5 O-type stars and 23 YSOs, while
clusters F and K have 30 or more YSOs and zero O-type stars.  These
differences point to different ages or different IMF slopes from one
cluster to the next, although when the full population of the South
Pillars is averaged together, we do not find compelling evidence for
an altered IMF (see \S 3.3).  This implies that small IMF differences
may be statistical fluctuations that are washed out in a larger
sample.

The relationship between clusters, YSOs, and O-type stars in the South
Pillars has important implications for the region.  Most of the O-type
stars in and around Cr~228 are usually assumed to be part of the same
cluster as Tr~16, just divided at visual wavelengths by a dark
obscuring dust lane, and that they therefore have the same age of
$\sim$3 Myr (e.g., Walborn 1995, 2009; Smith 2006).  This would make
Tr~16 much longer in the N-S direction than in the E-W direction.  If,
instead, the O stars in the South Pillars are associated with several
individual sub-clusters that are 10--20 pc from Tr~16, it suggests
that they were born in these sub-clusters along with the associated
YSOs and that they are therefore younger than Tr~16.  It also has
potential implications for the IMF in regions of second-generation
triggered star formation like the South Pillars, so detailed studies
of the stellar content, photometric properties, kinematics, and other
properties of these South Pillar clusters in Figure~\ref{fig:clusters}
may be quite fruitful.  If the WNH star HD~93131 was formed more
recently in the South Pillars, it has important implications for the
nature of WNH stars and the evolution of the most massive stars (see
Smith \& Conti 2008).

2.  While several pillars have one YSO, a few YSOs, or a small cluster
at the head of the pillar (many of which appear to drive HH jets;
Smith et al.\ 2010), it is also true that {\it most} of the star
clusters in Figure~\ref{fig:clusters} -- as well as the sub-clusters
of YSOs associated with them --- are spatially {\it offset} from the
heads of pillars by 1-5 pc.  The clusters seem to lag behind the
pillars in an outwardly propagating sense, so that ``behind'' means
closer to the center of the H~{\sc ii} region.  Striking examples are
clusters A, D, F, G, K, and N (cluster G appears tobe associated with
a small pillar to its east in Figs.~\ref{fig:clusters} and
\ref{fig:Tr16se}).  Both K and N seem to still be partially embedded,
and these are far from Tr~16 and associated with the largest pillars
or cloud fragments.  One would therefore expect them to be younger
than clusters A, D, F, and G, implying a north-to-south age gradient.

The main implication of this offset is that the pillars are transient
structures associated with the destruction of a cloud by feedback from
massive stars, and that the resulting star clusters are left behind in
their wake.  It is interesting to compare the structure of PAH
emission observed in some of the pillars to simulations of radiative
and shock destruction of clouds (e.g., Pittard et al.\ 2009; Marcolini
et al.\ 2005; Mizuta et al.\ 2006).  In those simulations, by the time
the pillar achieves the highly fragmented and filamentary structure
comparable to the observed structures for many pillars in Carina, the
head of the pillar has been pushed backward or evaporated.  As such,
the current position of the pillar head is offset from the initial
cloud position by an amount commensurate with the observed offset
between star clusters and pillars in Carina (compared to the relative
size of the pillar).  The expected cloud destruction time of several
$\times$10$^5$ yr is comparable to the ages of the Stage II YSOs in
the young custers. This offset between the young clusters and stars
has important implications for the question of triggered star
formation, and the relative kinematics of the stars and gas are
paramount (see \S 5).

3.  As we noted in \S 3.2, most of the YSOs are concentrated in a
large grouping that fills the interior cavity encircled by the PAH
emission from the South Pillars.  Following the previous point about
sub-clusters residing in the wakes of pillars, this large association
of YSOs in the South Pillars appears to have been built up over time,
as a hierarchical group contributed by many pillars and sub-clusters
rather than one monolithic cluster from a single cloud core.  The end
product of this is a large loose association of $\sim$20 O-type stars
and many lower-mass stars, although the star formation here is by no
means complete yet.

4. A corallary of the previous point is that there are also several
dust pillars that appear to harbor {\it no} YSOs or clusters at their
tip.  This is especially true in the West region, where no clusters or
tight groupings of YSOs could be identified.  One of the western
pillars that does have a clear YSO at its head is the likely driving
source of the HH~1010 jet (Smith et al.\ 2010; see
Figure~\ref{fig:colorWest}).  Overall, the spatial distribution of
stars and YSOs in the West region is more diffuse than in the South
Pillars, and less correlated with the PAH emission from pillars.  This
implies that, again, the YSOs are left behind in the wake of a
continuously propagating wave of star formation, but which is less
active and less fragmented than in the South Pillars.

\subsection{An Obscured Massive Star Cluster Near Tr~16}

The {\it Spitzer}/IRAC images of Carina have revealed several small
clusters or subclusters in the South Pillars, discussed in the
previous section.  One of these --- cluster G in
Figure~\ref{fig:clusters} and Table~\ref{tab:cluster} --- deserves
more detailed attention, because it contains an interesting collection
of massive stars near Tr~16, but it is hidden behind the dark V-shaped
obscuring dust lane that bisects the Carina Nebula, and so it has not
been recognized until recently.  Figure~\ref{fig:Tr16se} shows {\it
  Spitzer}/IRAC and visual-wavelength images of the same field,
centered on this cluster (the optical narrow-band images were obtained
with the MOSAIC camera on the CTIO 4m telescope; Smith et al.\ 2003).

Sanchawala et al.\ (2007) first drew attention to some of these
objects, noting a compact group of ten X-ray sources that had near-IR
$K$-band counterparts, which they referred to as Tr~16-SE because of
their proximity to Tr~16.  They noted that two of these X-ray sources
were new OB star candidates, while one that was relatively less
obscured was a known O4-type star (MJ 596 or IRS-43; Smith 1987; see
Fig.~\ref{fig:Tr16se}).  Aside from this star, the remaining members
were invisible at optical wavelengths.  Sanchawala et al.\ (2007)
suggested that this group of X-ray sources may represent a deeply
embedded group of very young stars, and that by virtue of their
location in a dark cloud at the edge of an H~{\sc ii} region, this
group may have formed recently via the collect-and-collapse mode of
triggered star formation.

Our new {\it Spitzer} images reveal a somewhat different picture for
this group of stars.  The distribution of PAH emission in IRAC images
suggests that the cluster resides in a local cavity that has already
been cleared around it.  A bright edge-on PDR is seen to the east,
including a sharp dust pillar (G287.75-0.66) that points to one member
of the cluster, but the diffuse emission ends abruptly before reaching
Tr~16-SE (G).  The stars show no enhancement of collective diffuse
emission between them as seen in other embedded clusters in the region
like the Treasure Chest or IRAS~10430 (see Fig.~\ref{fig:Tr16se}).
Thus, IRAC images suggest that rather than being deeply embedded {\it
  inside} the dark cloud, Tr~16-SE is simply projected {\it behind}
the foreground obscuring cloud at visual wavelengths.  It appears to
reside within an extension of the same cavity cleared by Tr~16.  While
this alters the characterization of Tr~16-SE as an embedded cluster
whose formation may have been triggered recently by the interaction
between Tr~16 and the dark cloud seen today, it may nevertheless be a
fairly young cluster that formed sometime after Tr~16.  There are
$\sim$30 objects classified as YSOs in this cluster
(Fig.~\ref{fig:ysos}).  However, they are all classified as either
ambiguous (cyan) or stage II (yellow) in Figure~\ref{fig:stage}, which
implies evolution beyond the deeply embedded phase.  It suggests a
more mature group of stars with an age of $\sim$1 Myr, rather than an
extremely young cluster with an age of $\sim$10$^5$ yr as is seen in
the Treasure Chest (Smith et al.\ 2005).

%%%%%%%%%%%%%%%%%%%%%%%%% FIGURE 15 - cartoon  %%%%%%%%%%
\begin{figure*}\begin{center}
%\epsscale{0.99}
%\includegraphics[width=5.8in]{../FIGS/pillarsketch.eps}
\includegraphics[width=5.8in]{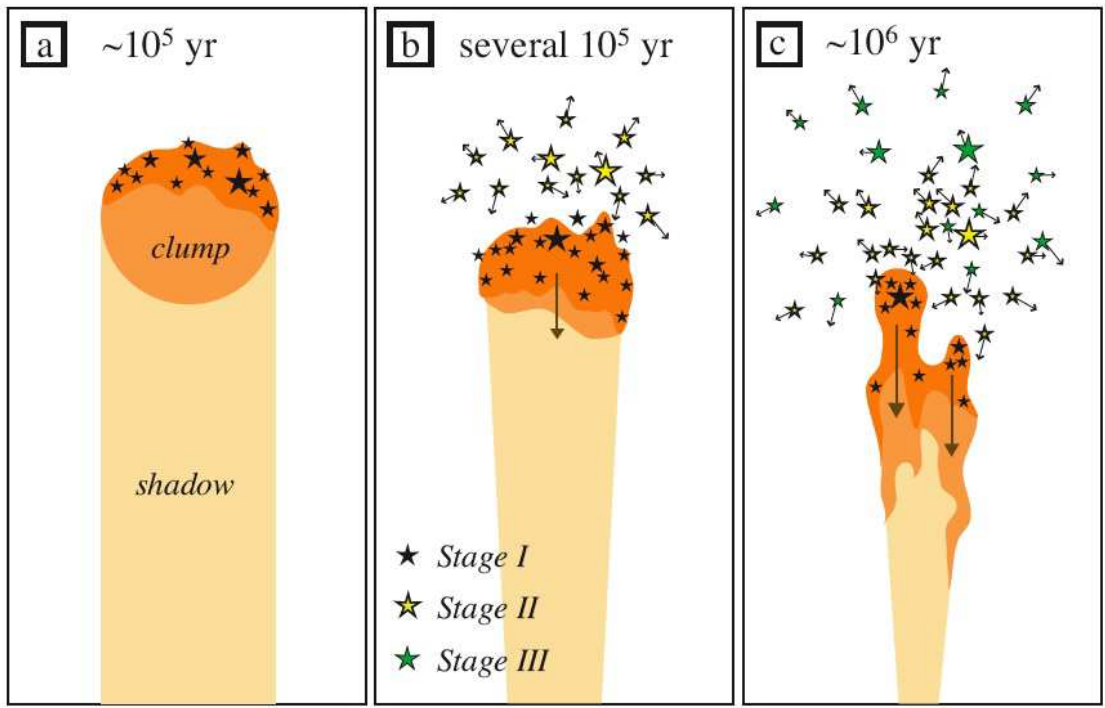}
\end{center}
\caption{Cartoon depicting feedback-induced star formation in the
  evolution of a single dust pillar.  The primary external source of
  winds and UV radiation (i.e. the Tr 14 and 16 star clusters in the
  case of Carina) is located off the top of the page.  Panels a, b,
  and c represent snapshots of three sequential phases as the cloud is
  destroyed.  Panel (a) depicts the beggining of the process, when
  feedback has swept back low density regions between clumps and first
  begins to compress the main clump, triggering the formation of new
  stars (Stage I protostars; black stars) on a timescale of
  $\sim$10$^5$ yr. Panel (b) shows a subsequent phase after a few
  10$^5$ yr.  The clump gas has been accelerated downward (Oort \&
  Spitzer 1955; Elmegreen \& Lada 1977), away from the first
  generation OB stars, and new stars are forming as it continues to be
  compressed (black stars).  After a few 10$^5$ yr, the Stage I
  protostars formed in Panel (a) are no longer embedded and are in the
  process of losing their disks, so they are now Stage II YSOs (yellow
  stars).  Because of random motions of the stars, this group of stars
  begins to disperse after the surrounding cloud has been removed.
  Panel (c) shows an even later phase after about 10$^6$ yr.  The
  remnants of the initial clump have been swept back from the initial
  position of the clump and now form a highly fragmented dust pillar
  that is forming new Stage I protostars.  The stars that formed in
  panel (b) are now Stage II YSOs, dispersing somewhat due to their
  random motions.  The stars that formed in Panel (a) have lost much
  of their disk material after $\sim$10$^6$ yr, appearing as Stage~III
  YSOs (green stars), and have spatially dispersed even more.  The
  random motions of the stars after gas removal causes the various
  stellar generations to mix across a large region above the dust
  pillar.  When multiple pillars are involved as in Carina, these
  stars are subsumed into a larger association of stars.  A small
  cluster of YSOs may or may not be seen to be associated with a given
  pillar, depending on the veracity of star formation in a previous
  phase.}\label{fig:sketch}
\end{figure*}
%%%%%%%%%%%%%%%%%%%%%%%%%%%%%%%%%%%%%%%%%%%%%%%%%%%%%%%%%%%%%%%%%%%%%%

The earliest type star known in the region is MJ-596 (or IRS-43),
which has a spectral type of O4 V (Smith 1967), and which Kukarkin et
al.\ (1981) note is a possible eclipsing binary with a somewhat
different spectral type O5.5 V.  However, several other stars in the
group, while very faint or undetected at visual wavelengths, are
brighter than MJ~596 in IRAC images with colors of mildly reddened
stellar photospheres.  Sanchawala et al.\ (2007) also noted possible
OB candidates based on X-ray emission and IR colors.  Thus, Tr~16-SE
may be a fairly substantial sub-cluster of OB stars that has not yet
been recognized.  It would be interesting to perform a more detailed
analysis of the age and stellar content of this group, and to obtain
near-IR spectra of all bright members for spectral typing (e.g.,
Hanson et al.\ 1996).

\section{The Steady March of Propagating Star Formation and the
  Buildup of an OB Association}

Together, the {\it Spitzer}/IRAC results for Carina suggest an
interesting scenario where feedback from a first generation of massive
OB stars drives an outwardly propagating wave of star formation, which
simultaneously destroys the initial molecular cloud and continually
forms new generations of stars.  Our results suggest that this may be
a steady and ongoing process, rather than a simple 2-stage process.
As an initially clumpy cloud is sculpted into a series of dust
pillars, feedback also accelerates those dust pillars outward.  As
such, the wave of star formation propagating out through the pillars
leaves behind a wake of newly formed stars in the void between the
first generation OB stars and the pillars.

%%%%%%%%%%%%%%%

Figure~\ref{fig:sketch} is a cartoon depicting the process of
triggered star formation within dust pillars. In the first stage
represented here (Figure~\ref{fig:sketch}a), feedback from a first
generation of massive stars (located off the top of the page)
interacts with a clump in a molecular cloud.  Regions of lower density
outside the clump are ionized and swept back, leaving a clump with an
ionization shadow below it (e.g., Williams et al.\ 2001). The
ionization/shock front impinging on the cloud compresses its surface,
and some regions in this compressed gas are unstable to self gravity
and collapse to form new stars.  These are the relatively rare
(i.e. short-lived) Class 0/Class I YSOs that are embedded in dust
pillars and are seen to drive HH jet outflows from the heads of
pillars and cometary clouds in Carina (Smith et al.\ 2010).  In
optical images, the structure at this stage will likely resemble a
large dark pillar such as the Giant Pillar in Carina (see
Figure~\ref{fig:colorSP}) or other features that still appear
connected to their parent molecular clouds.

Figure~\ref{fig:sketch}b depicts a more advanced stage, when the
pillar has been eroded and accelerated outward, and is now forming
another generation of YSOs.  The stars that formed in
Figure~\ref{fig:sketch}a are now Stage II YSOs (yellow), and are
beginning to disperse due to their own random motions (represented by
small arrows in the cartoon) after gas removal.  This stage may be
representative of moderate-sized pillars in Carina with active star
formation, such as the one that harbors the Treasure Chest cluster
(Smith et al.\ 2005).

Finally, Figure~\ref{fig:sketch}c shows an even more advanced stage of
the process.  The pillar is now a more filamentary and fragmented
remnant of the original clump that has been largely destroyed by
instabilities (e.g., Pittard et al.\ 2009).  Previous generations
formed in Figures~\ref{fig:sketch}a (green stars) and b (yellow stars)
are now dispersing further, spreading out in the wake of the
accelerated dust pillar.  Figure~\ref{fig:sketch}c portrays the
interesting consequence that previous generations of YSOs formed in
the pillar do not remain in the pillar and can be mixed with one
another due to their own velocity dispersions after gas removal, over
an area considerably larger than the observed size of the pillar head.
This makes it difficult to pick out a coherent sub-cluster at a
specific age or to make a conclusive association with that pillar.
The degree to which newly born stars inherit the bulk velocity of
accelerated gas out of which they form is an interesting goal for
future work comparing numerical simulations to observations of stellar
and gas kinematics.  For example, we might expect successive
generations to have slightly different stellar kinematics, depending
on the speed of the accelerated gas.  Figure~\ref{fig:sketch} is of
course idealized, with just a single pillar, starting at a given time
in Figure~\ref{fig:sketch}a with no previous star formation aside from
the massive stars that drive the feedback; the true picture is likely
to be complicated.

Now suppose that there is an ensemble of these pillars, each of which
may have originated at a different distance from the source of
feedback (and hence is at a different stage of evolution) and with a
different initial mass of the clump.  One can imagine that coherent
sub-clusters of stars could be rather difficult to attribute to any
individual pillar, due to confusion caused by the dispersal of the
YSOs left behind by the outwardly advancing pillar.  Furthermore, an
ensemble of pillars can have a range of orientations projected on the
sky with some pillars behind others or overlapping, so the spatial
arrangement of stars and pillars can get quite complicated.
Nevertheless, we do observe the general trend that the large
association of YSOs is generally outlined by dust pillars, and we do
see some examples of sub-clusters of YSOs associated with nearby
pillars.  The YSOs have a much lower space density outside the region
bounded by the dust pillars, indicating that this star formation has
indeed been triggered by feedback.  We propose, therefore, that this
general picture can explain the observed properties of the South
Pillars in Carina, and is probably the dominant mode of ongoing star
formation in more distant evolving giant H~{\sc ii} regions such as
30~Dor and NGC~604.

%%%%%%%%%

The stars forming in this propagating wave are not necessarily all low
mass stars.  Indeed, the region of the South Pillars contains roughly
10--20 O-type stars (it is not clear if they all formed there), one of
which is a very massive WNH star (HD~93131).  It has generally been
assumed that the O stars in Cr228 make up an extended part of Tr~16,
but the results here suggest that at least some of the O-star
components in Cr228 were born in the South Pillars about 1--2 Myr
after Tr~16.  If true, this has important implications for the IMF of
this propagating star formation (i.e. it forms both massive and
low-mass stars; see \S 3).  These O-type stars may profoundly affect
the subsequent shaping of the pillars because they are much closer to
the gas than the original first generation O-type stars.  Examining
Figure~\ref{fig:sketch}c, for example, one can easily imagine that a
newly born massive star located off the axis of the pillar (because it
formed in a neighboring pillar, or because it drifted there due to its
own random motion) could bend the apparent elongation axis of the
pillar.  This is especially true for smaller and more evolved
filamentary pillars or cometary clouds.  Indeed, upon examining the
orientations of dust pillar axes in the South Pillars of Carina
(Figure~\ref{fig:pillars}), one sees that most of the large pillars
point toward $\eta$ Carinae and the Tr~14 and Tr~16 clusters, whereas
many of the smaller pillars and cometary clouds point in other
directions.  One can even see cases where a large pillar has an axis
that points toward the main clusters, whereas some of its own
sub-components point elsewhere, as one might infer from
Figure~\ref{fig:sketch}.

%%%%%%%%%%%%%%%

In \S 3 we identified roughly 1000 YSOs with ages of 0.1--1 Myr, but
because of incompleteness due to various effects --- exclusion due to
our detection criteria, extended emission, background contamination,
etc., regions not included in our IRAC map, and especially due to
detection limits that exclude most YSOs below 2--3 $M_{\odot}$ --- we
estimated that the true number of YSOs in this age range is closer to
5$\times$10$^3$ or more.  These are located within a larger region
that is roughly 3 Myr old and has formed a total of roughly
(5--8)$\times$10$^4$ stars (Smith 2006; Smith \& Brooks 2007).  Very
roughly speaking, the size of the younger population seen currently
seems consistent with a relatively constant rate of star formation
over the 3 Myr age of the Carina Nebula.

Stars in Tr 16 are usually assumed to be $\sim$3 Myr old, whereas Tr14
is probably somewhat younger at roughly 2 Myr (Walborn 1995; Smith
2006).  The various sub-clusters of YSOs in the South Pillars
discussed here constitute another major component of the stellar
content.  It therefore seems likely that the Carina Nebula complex is
an OB association that is built up steadily over time, throughout the
age of the region and continuing today.  This departs from the
prevailing view of the region, which has generally assumed a first
generation of OB stars formed in an isolated event (Tr~14 and 16),
followed by relatively low-level ``percolating'' star formation
induced by feedback (see review by Smith \& Brooks 2007).  This
implies that the current mode of star formation within the pillars of
Carina is an important mechanism for the continual buildup of OB
associations.

%%%%%%%%%%%%%%%%%%%%%%%

The region contains a total mass in stars of (4--6)$\times$10$^{4}$
$M_{\odot}$ (Smith 2006; Smith \& Brooks 2007), and is currently
spread across a region of $\sim$30 pc in size, so the escape velocity
from the region is only $\sim$4 km s$^{-1}$.  After the gas has been
evacuated, the stars are therefore likely to be unbound due to their
own random motions, eventually constituting a loose OB association
rather than a bound cluster.  (Subcomponents such as Tr~14, however,
may remain bound as open clusters for a significant time.)  It
therefore seems clear that in Carina we are witnessing the birth and
continual buildup of an OB association over an extended period of
several Myr.  Smith (2006) noted, for example, that the stellar
content of the Carina Nebula is comparable to that of the cluster in
NGC~3603, although in Carina it is spread among several sub-clusters
across 10--30 pc.

We conjecture that the process of propagating star formation across a
cloud may be an important key in forming an unbound OB association
rather than a bound cluster, since stars are forming in gas that is in
the process of being accelerated outward by feedback.  If this is
common, one would therefore expect significant age spreads in OB
associations, of order a few Myr.  If massive stars form in the
dispersed population and they inherit the outward motion of the
accelerated pillars, then some substantial fraction of the massive
stars in an OB association may migrate quite far from their birth
site.  Expanding outward at $\sim$10 km s$^{-1}$, a massive star can
move 100 pc in 10 Myr.  Viewed after several Myr near the end of their
main-sequence lifetime, this may produce a population of late O-type
stars that will appear to be in relative isolation, even though they
formed at the periphery of a giant H~{\sc ii} region.  When they
finally explode as a SN, they will not be associated with an obvious
H~{\sc ii} region or star cluster.  In Carina, this may apply to
$\sim$20 of the 70 O-type stars in the whole region.

%%%%%%%%%%%%%%%%%%%%%%%%

In future observations, it will be interesting to directly measure the
kinematics of Carina's stars and gas independently, to confirm the
scenario of star formation being triggered as gas in the pillars is
accelerated by feedback, as we depict in Figure~\ref{fig:sketch}, and
to test the degree to which the association is gravitationally bound.
Comparison to numerical simulations of feedback effects on clouds that
include self gravity will be key to this effort.  Also, a comparison
of our YSO sample to the sample of young stars traced by their X-ray
emission is an important next step to investigate the population of
stars that are losing their disks.

%%%%%%%%%%%%%%%%%%%%%%%%%%%%%%%%%%%%%%%%%%%%%%%%%%%%%%%%%%%%%%%%%%%%
%%%%%%%%%%%%%%%%%%%%%%%%%%%%%%%%%%%%%%%%%%%%%%%%%%%%%%%%%%%%%%%%%%%%
\section{Conclusions}

We have analyzed multiwavelength images of two regions in the Carina
Nebula obtained with the IRAC camera onboard {\it Spitzer}, including
both point sources and the diffuse emission from dust and PAHs.  We
have conducted photometry and produced a point source catalog merged
with data from the 2MASS point source catalog, and we have used this
to produce a list of highly reliable YSOs selected on the basis of
fits to their SEDs.  Here we provide a brief list of several
conclusions from this work.

1.  We provide a merged IRAC+2MASS catalog of over 44,000 point
sources detected in at least 4 of 7 filters.  Most of these are
foreground and background sources.  We identify 909 YSOs, selected
based on fits to their SEDs.  We note that this is a severe
underestimate of the true number of YSOs for several reasons,
including our rigorous selection criteria for point sources.

2.  Our YSO sample is also an underestimate of the true number of YSOs
because it misses faint sources due to sensitivity at stellar masses
below 2--3 $M_{\odot}$.  Correcting for these missing sources using an
Orion-like IMF (Muench et al.\ 2002), we find that there should be
more than 5,000 YSOs in the South Pillar region of Carina.
Considering that even this is an underestimate, we find it likely that
star formation has continued at a relatively constant level over the 3
Myr lifetime of the region.  The current star formation occurring in
dust pillars may therefore represent an important mode for the
continual buildup of large OB associations.

3.  Comparing the $K$-band and 3.6 $\mu$m luminosity functions (LFs)
of our YSO sample to the LFs from Orion, we find no compelling
evidence that the IMF is different among the collective generation of
stars currently forming in Carina.  There are slight deficits in
Carina at the bright end of the LFs, which are subject to low number
statistics from the Orion sample, but we speculate that the apparent
deficit of more luminous YSO sources in Carina compared to Orion may
result if more massive stars shed their own disks more quickly than
lower mass stars.  The 10--20 O-type stars that are spatially
associated with the YSOs in the South Pillars probably formed along
with them, and none of these have dusty disks.

4.  A corollary to the previous point is that while the aggregate
population shows no convincing deviations from a standard IMF, there
do appear to be possible fluctuations in the IMF from one sub-cluster
to the next.  This is indicated by the apparent ratio of the numbers
of O-type stars to associated YSOs, although this may also be an age
effect.

5.  We detect surprisingly few of the so-called extended green objects
(EGOs; thought to be molecular outflows), given the large number of
YSOs and the known outflow activity traced by optical HH jets (Smith
et al.\ 2010).  We attribute this lack of EGO sources to
photodissociation of the molecular outflows by the UV radiation field
in Carina and to the added difficulty of detecting excess Band 2 flux
amid the bright background emission in Carina.

6.  A population of extended red objects (EROs) exhibit diffuse
emission that is not consistent with the colors expected for PAH
emission.  The EROs are found to be associated with late O-type or
early B-type stars, and several show a bow-shock morphology with the
apex of the shock pointing inward to the first generation massive
stars.  We suggest that these EROs represent thermal dust emission in
shocks at the interface between stellar winds and dense
photoevaporative flows in the South Pillars, akin to the similar
structures seen in M~17 (Povich et al.\ 2008).

7.  Judging by qualitative aspects of the observed structures of
pillars seen in PAH emission, their filamentary structure is
consistent with shocked or photoablated clouds in advanced stages of
destruction.  The smaller pillars have probably been accelerated
outward from their initial positions.

8.  We have analyzed the directions of the pillar axes for the
ensemble of pillars in Carina.  While large pillars generally point
inward to the first generation O-type stars in Tr~14 and Tr~16, many
of the smaller pillars and cometary clouds point in other directions.
We suggest that as a cloud is sculpted into the shape of a pillar by
the first generation stars, it may also succumb to the influence of
local O-type stars that were born in the second generation or that
moved into its vicinity, effectively bending the pillar to a different
orientation.  Local O-type stars may also be more influential when the
most massive stars like $\eta$ Car reach the ends of their lives and
their ionizing flux drops.

9.  The relative spatial distribution of YSOs and dust pillars reveals
that while several Stage~I YSOs are indeed located within the heads of
dust pillars, many more Stage~I, II and III YSOs are scattered outside
of pillars.  The YSOs show some small-scale clumping, but in general
they form a large association of YSOs occupying a cavity that is
bounded by pillars.  In several cases, smaller sub-clusters of YSOs
are found just interior to the current locations of the heads of dust
pillars.

10. We draw attention to one subcluster of stars, which was shown
previously to have several X-ray sources (Sanchawala et al.\ 2007).
We argue that this cluster (Tr 16 SE) is not an embedded cluster, but
is instead a young cluster that is obscured behind (not within) the
dark dust lane that bisects the Carina nebula, with a likely age of
$\sim$1--2 Myr.

11. We propose a scenario where pillars are transient features in a
continually outwardly propagating wave of star formation.  As the dust
pillars are accelerated outward and destroyed by feedback from
first-generation massive stars, they also form new stars in the
process and leave behind a wake of YSOs.  The YSOs formed over time by
an ensemble of pillars are subsumed into a young OB association formed
over a span of $\sim$10$^6$ yr, but the youngest ones still show some
hierarchical sub-clustering and spatial association near the heads of
pillars.

12.  The current star formation rate compared to the existing
population of massive stars in the Carina Nebula is roughly consistent
with a relatively {\it constant} rate of star formation averaged over
$\sim$3 Myr.  Roughly 10--20 of the region's 70 O stars are closely
associated with gas and dust pillars in the South Pillar region, and
appear to have formed there recently (in the last $\sim$1 Myr, rather
than 3 Myr ago like Tr16), based on the Stage II and III YSOs that
they are associated with).  This is consistent with our finding above
that there is no compelling evidence for a substantially altered IMF.
Thus, the current star formation in the Pillars of Carina is likely to
represent an important mode of star formation for the gradual buildup
of an OB association.

13.  We find that the fledgling OB association is likely to be
unbound.  When all star formation in the region has ceased and the gas
is removed, we will likely be left with a dispersing OB association
where the massive stars have an age spread of 3--4 Myr.  There will be
a ``halo'' of stars spread across $\sim$100 pc, surrounding a pair of
central clusters (Tr~14 and 16) that will be marginally bound;
altogether it may appear much like h and $\chi$ Persei (Currie et al.\
2009).  The imminent onset of the supernova driven phase may impact
this picture, however, as discussed by Smith \& Brooks (2007). We
suggest that the nature as an unbound association is likely related to
its formation mechanism in an outwardly propagating wave distrubuted
over space rather than in a single compact cluster.

14.  Feedback-driven star formation may result in a substantial
fraction of the massive-star content being spatially dispersed, spread
across $\sim$100 pc after several Myr.  Migrating from their birth
site, these massive stars may falsely appear to have formed in
isolation, and their SNe will not necessarily be coincident with an
obvious H~{\sc ii} region or star cluster. More detailed consideration
of these points will be discussed in a future paper.

%\acknowledgments \footnotesize
\smallskip\smallskip\smallskip\smallskip
\noindent {\bf ACKNOWLEDGMENTS}
\smallskip
\scriptsize

This work was based on observations made with the Spitzer Space
Telescope, which is operated by the Jet Propulsion Laboratory,
California Institute of Technology under a contract with NASA. Support
for this work was provided by NASA through awards issued by
JPL/Caltech as part of GO programs 3420 and 20452.%, and 30848 - MIPS.
M.S.P.\ is supported by an NSF Astronomy \& Astrophysics Fellowship
under award AST-0901646.  B.A.W.\ was supported by NASA through the
Spitzer Space Telescope Theoretical Research Programs, through a
contract issued by the JPL/Caltech under a contract with NASA.
R.D.G.\ was supported by NASA through contracts No.\ 1256406 and
1215746 issued by JPL/Caltech to the University of Minnesota.

% REFERENCES

\end{document}